\documentclass[fleqn,usenatbib]{mnras}

\usepackage{newtxtext,newtxmath}

\usepackage[T1]{fontenc}

\DeclareRobustCommand{\VAN}[3]{#2}
\let\VANthebibliography\thebibliography
\def\thebibliography{\DeclareRobustCommand{\VAN}[3]{##3}\VANthebibliography}

\usepackage{graphicx}	%
\usepackage{amsmath}	%
\usepackage{multicol}        %
\usepackage{bm}		%
\usepackage{pdflscape}	%
\usepackage{newtxtext,newtxmath}
\usepackage[T1]{fontenc}
\usepackage{ae,aecompl}

\title[Alignments in the cosmic web]{On the alignment of haloes, filaments and magnetic fields in the simulated cosmic web}

\author[S. Banfi et al.]{S. Banfi$^{1,2}$\thanks{Contact e-mail: \href{serena.banfi2@unibo.it}{serena.banfi2@unibo.it}}, F. Vazza$^{1,2,3}$, C. Gheller$^{2}$\\
$^{1}$Dipartimento di Fisica e Astronomia, Universit\`a di Bologna, Via Gobetti 92/3, 40129, Bologna, Italy\\
$^{2}$Institute of Radioastronomy - INAF, Via Gobetti 101, 40129 Bologna, Italy\\
$^{3}$Hamburger Sternwarte, Gojenbergsweg 112, 21029 Hamburg, Germany}

\date{Accepted XXX. Received YYY; in original form ZZZ}

\pubyear{2020}

\begin{document}
\label{firstpage}
\pagerange{\pageref{firstpage}--\pageref{lastpage}}
\maketitle

\begin{abstract}
The continuous flow of gas and dark matter across scales in the cosmic web can generate correlated dynamical properties of haloes and filaments (and the magnetic fields they contain). With this work, we study the halo spin properties and orientation with respect to filaments, and the morphology of the magnetic field around these objects,  for haloes with masses in the range $\sim 10^8 - 10^{14}\ \mathrm{M_{\odot}}$ and filaments up to $\sim 8\ \mathrm{Mpc}$ long. Furthermore, we study how these properties vary in presence, or lack thereof, of different (astro)physical processes and with different magnetic initial conditions. We perform cosmological magnetohydrodynamical simulations with the Eulerian code \textsc{Enzo} and we develop a simple and robust algorithm  to study the filamentary connectivity of haloes in three dimensions.  We investigate the morphological and magnetic properties and focus on the alignment of the magnetic field along filaments: our analysis suggests that the degree of this alignment is partially dependent on the physical processes involved, as well as on magnetic initial conditions. We discuss the contribution of this effect on a potential attempt to detect the magnetic field surrounding these objects: we find that it introduces a bias in the estimation of the magnetic field from Faraday rotation measure techniques. Specifically, given the strong tendency we find for extragalactic magnetic fields to align with the filaments axis, the value of the magnetic field can be underestimated by a factor $\sim 3$, because this effect contributes to making the line-of-sight magnetic field (for filaments in the plane of the sky) much smaller than the total one.
\end{abstract}




\section{Introduction}
The evolution of the Universe by hierarchical clustering has led to the assembly of different structures, characterised by being either underdense or overdense to different extents, like voids, walls, filaments and haloes. Connected together, all of these elements constitute the so called cosmic web, a network of dark and baryonic matter, which links all kinds of structures in a distinctive, intricate arrangement \citep{1996Natur.380..603B}. The pattern of the cosmic web is the manifestation of the tidal field arisen from the inhomogeneous distribution of matter as a result of anisotropic gravitational collapse \citep{1970SvA....13..608Z}. According to the Zeldovich approximation, this collapse initially induces the contraction of matter into walls, then filaments, and eventually into fully collapsed entities \citep{1982GApFD..20..111A,1984SvA....28..491S,1989RvMP...61..185S,1989MNRAS.236..385G,2014MNRAS.437.3442H}. The connectivity of these components can be explained by Bond's theory \citep{1996Natur.380..603B} as an effect of the tidal shear, which generates the quadrupolar mass distribution leading to a typical cluster-filament-cluster configuration. Hence, filamentary structures are formed in environments where shear stresses are effective between the overdense matter and voids, thus dragging the gas along the spine of the filament.

The tidal field is also believed to be responsible for the acquisition of angular momentum by these structures, following the Tidal Torque Theory (TTT) \citep{hoyle1949problems,1969ApJ...155..393P,1970Afz.....6..581D,1984ApJ...286...38W}, thus linking the rotation properties of haloes to their surroundings' density distribution. According to TTT, the halo spin should initially be correlated with the principal axes of the local tidal tensor, and in  particular be perpendicular to the hosting filament orientation. However, many studies showed that there actually is a transition mass below which spins are mostly parallel to the filament, and above which the preferential arrangement is perpendicular: the value of the so called \textit{spin flip mass} is reported to span from $\sim 0.5$ to $\sim 5 \times 10^{12}\ h^{-1}\ \mathrm{M_{\odot}}$ in different works \citep[e.g.][]{2007A&A...474..315A,2007MNRAS.375..489H,2010MNRAS.405..274H,2012MNRAS.427.3320C,2013MNRAS.428.2489L,2013ApJ...762...72T,2014MNRAS.444.1453D,2014MNRAS.443.1090F,2017MNRAS.468L.123W}. This effect is believed to be due to a non-linear phase of TTT, involving mergers or accretion of substructures \citep{2014MNRAS.445L..46W, 2012MNRAS.420.3324B,2016MNRAS.461.1338B}.

This trend is supported by galaxy observations: spin properties of galaxies can be obtained from their rotation curves, with some assumptions on galaxy morphological properties \citep[e.g.][]{2006MNRAS.368..351H}. Observational studies vastly confirm that spins of spiral galaxies (associated to less massive haloes) are mostly parallel to the host filament, while elliptical galaxies (associated to more massive haloes) tend to spin along the direction normal to the filament \citep[e.g.][]{2013MNRAS.428.1827T,2010MNRAS.402.1807C,2013ApJ...775L..42T,2013ApJ...779..160Z,2016MNRAS.457..695P,2017A&A...599A..31H}.

Among the various large-scale fields that can develop a relevant correlation across scales, are also extragalactic magnetic fields \citep[e.g.][]{ry08}, as we preliminary explored in  \citet{2020MNRAS.496.3648B}. The formation dynamics of the cosmic web is indeed also found to affect the large-scale topology of magnetic fields, for a variety of possible seeding scenarios. In particular, in \citet{2020MNRAS.496.3648B} we studied the angle formed by the propagation direction of cosmic shocks and the up-stream magnetic field (i.e. obliquity), which is believed to be a crucial parameter for cosmic-ray acceleration by shocks \citep[e.g.][and references therein]{Bykov2019}: with cosmological simulations we measured that magnetic field lines tend to align to filaments both inside and outside the filament, following the flow direction on the gas, as a consequence of the velocity shear. This effect, which was found to apply to several variations of primordial scenarios of magnetic fields \citep[][]{va21} as well as to variations of astrophysical seeding scenarios, albeit in a less prominent way \citep[][]{2020MNRAS.496.3648B},  impacts on obliquity and therefore on cosmic-ray acceleration. 

The trend outlined above, on one hand being extremely relevant for the study of cosmic-ray acceleration and cosmic magnetism, is also very challenging to detect in observations. In this new work, we seek a way to assess the likely topology of magnetic fields around large-scale structures, based on the local properties of filaments and of the haloes they contain. In practice, we want to determine whether morphological and dynamical properties of the cosmic web components are sufficient to adequately predict the characteristics of the magnetic field: in particular, in this work we shall look for a correlation between haloes' angular momenta, filament orientation and magnetic field topology. Since in principle such properties may vary for different magnetic models, we also investigate different scenarios for the origin of extragalactic magnetic fields, which is believed to be either primordial or astrophysical: this introduces some degree of uncertainty on its topology, especially around structures like galaxy clusters and filaments \citep{2016RPPh...79g6901S}.

This paper is structured as follows. In Section \ref{sec:met}, we describe the computational setup for the simulations and we outline the network reconstruction method. In Section \ref{sec:res}, we present our results for spin-filament and filament-magnetic field alignment. In Section
\ref{sec:disc}, we describe the possible implications of our results on observations, as well as the numerical limitations encountered in our analysis. Finally, Section \ref{sec:conc} contains a brief summary and conclusions.

\section{Methods}
\label{sec:met}
\subsection{Simulations}
The simulations of this work are performed with the Eulerian cosmological magnetohydrodynamical code \textsc{Enzo} \citep[][]{enzo14}, which couples an N-body particle-mesh solver for dark matter \citep{he88} with an adaptive mesh refinement method  for the baryonic matter \citep{bc89}. We used a piecewise linear method \citep{1985JCoPh..59..264C} with Dedner cleaning MHD solver \citep{ded02} and time integration based on the total variation diminishing second-order Runge-Kutta scheme \citep{1988JCoPh..77..439S}.
In this work, we present the analysis of simulations of different volumes, resolutions and scenarios of the origin of magnetic fields. In particular, we analyze two sets of simulations, which will be referred to as \textit{``Roger''} and \textit{``Chronos''} (see Table \ref{tab}).
While the first is intended to test the resolution-dependent trends in the properties of the components cosmic web (for a relatively small cosmic volume), the second is designed to allow us to monitor how the properties of large-scale magnetic fields are related to the orientation of cosmic filaments, for a few relevant variations of the assumed origin scenario of cosmic magnetism.
For both sets, the cosmological parameters were chosen accordingly to a $\Lambda$CDM cosmology: $H_0=67.8\ \mathrm{km\ s^{-1}\ Mpc^{-1}}$, $\Omega_{\mathrm{b}}=0.0468$, $\Omega_{\mathrm{m}}=0.308$, $\Omega_{\mathrm{\Lambda}}=0.692$ and $\sigma_8=0.815$ \citep[][]{2016A&A...594A..13P}.

\subsubsection{Roger}
We first simulated a small volume of $\approx(19\ \mathrm{Mpc})^3$ 
(comoving) sampled with a static grid of $512^3$ cells, with the following characteristics:
\begin{enumerate}
 \item \textit{``NR''}: non-radiative run with a primordial uniform volume-filling comoving magnetic field $B_0=0.1\ \mathrm{nG}$ at the beginning of the simulation;
 \item \textit{``cool''}: radiative run including cooling, with a primordial uniform volume-filling comoving magnetic field $B_0=0.1\ \mathrm{nG}$ at the beginning of the simulation.
\end{enumerate}
Two additional simulations were run, similar to NR, in which the same volume is sampled by $256^3$ and $128^3$ cells, as a resolution test (see Section \ref{app1}).
The mass resolution for dark matter in the three Roger simulations is $6.3 \cdot 10^{8}\ \mathrm{M_{\odot}}$, $7.9 \cdot 10^7\ \mathrm{M_{\odot}}$ and $9.9 \cdot 10^6\ \mathrm{M_{\odot}}$ for the $128^3$, $256^3$ and $512^3$ runs respectively.

\subsubsection{Chronos}
\label{sec:detchron}
We simulated a volume of $\approx(84\ \mathrm{Mpc})^3$ (comoving) sampled with a static grid of $1024^3$ cells. We selected four runs taken from a larger dataset{\footnote{\textit{``Chronos++''} suite: \url{http://cosmosimfrazza.myfreesites.net/the\_magnetic\_cosmic\_web}}} provided by \citet{va17cqg}, which covers different possible scenarios for the origin and evolution of cosmic magnetic fields. The models chosen for this analysis are characterised by the following magnetic properties:
\begin{enumerate}
 \item \textit{``baseline''}: non-radiative run with a primordial uniform volume-filling comoving magnetic field $B_0=1\ \mathrm{nG}$ at the beginning of the simulation;
  \item \textit{``Z''}: non-radiative run with a primordial magnetic field oriented perpendicularly to the velocity vector, as in \citet{va17cqg}, accordingly to the Zeldovich approximation \citep[e.g.][]{do08}, in such a way to prevent $\nabla \cdot \mathbfit{B}$ from deviating from $\approx 0$; 
 \item \textit{``DYN5''}: non-radiative run with an initial seed magnetic field of $B_0 = 10^{-9}\ \mathrm{nG}$ (comoving) and sub-grid dynamo magnetic field amplification computed at run-time, which allows to estimate the hypothetical maximum contribution of dynamo in low density environments \citep[see][]{ry08}, where it would be lost due to finite resolution effects \citep[see][for more details]{va17cqg};
 \item \textit{``CSFBH2''}: radiative run with an initial seed magnetic field $B_0=10^{-10}\ \mathrm{nG}$ (comoving) including gas cooling, chemistry, star formation, thermal/magnetic feedback from stellar activity and active galactic nuclei (AGN). Supermassive black hole (SMBH) particles with a mass of $M_{\mathrm{BH,0}}=10^4\ \mathrm{M_{\odot}}$ are inserted at $z=4$ at the centre of massive haloes \citep[][]{2011ApJ...738...54K} and start accreting matter according to the Bondi-Hoyle formula (with an accretion rate of $\sim 0.01\ \mathrm{M_{\odot}/yr}$ and a ``boost'' factor of $\alpha_{\mathrm{Bondi}} =1000$, to compensate for gas clumping unresolved by the simulation). Star forming particles are generated according to \citet{2003ApJ...590L...1K}, which includes the contribution of stellar winds to the thermal feedback. Magnetic feedback from bipolar jets is introduced into the system, with efficiencies $\epsilon_{\mathrm{SF,b}}=10 \ \%$ and $\epsilon_{\mathrm{BH,b}}=1 \ \%$ for star formation and SMBH respectively \citep[see][for more details]{va17cqg}.
\end{enumerate}
The mass resolution for dark matter in all Chronos runs is  $8.1 \cdot 10^{7}\ \mathrm{M_{\odot}}$.

\begin{table*}
\begin{tabular}{|c|c|c|c|c|c|c|c|}
\hline
 \textbf{Name} & \textbf{Set} & \textbf{Details} & \textbf{B$_0$} & \textbf{Comoving volume} & \textbf{Cells} & \textbf{Spatial resolution} & \textbf{Dark matter resolution}\\
   & &  & $\mathrm{[nG]}$ & $\mathrm{[Mpc^3]}$ & & $\mathrm{[kpc/cell]}$ &$\mathrm{[M_{\odot}}]$\\ \hline
\textit{NR} & Roger & non-radiative & $0.1$ & $19^3$ & $512^3$ & $37$  & $9.9 \cdot 10^6$\\ \hline
\textit{NR$_{256}$} & Roger & non-radiative & $0.1$ & $19^3$ & $256^3$ & $74$ & $7.9 \cdot 10^7$ \\ \hline
\textit{NR$_{128}$} & Roger & non-radiative & $0.1$ & $19^3$ & $128^3$ & $148$ & $6.3 \cdot 10^{8}$ \\ \hline
\textit{cool} & Roger & cooling & $0.1$ & $19^3$ & $512^3$ & $37$ & $9.9 \cdot 10^6$\\ \hline
\textit{baseline} & Chronos & non-radiative & $1$ & $84^3$ & $1024^3$ & $82$ & $8.1 \cdot 10^{7}$ \\ \hline
\textit{Z} & Chronos & tangled magnetic field & $1$ & $84^3$ & $1024^3$ & $82$ & $8.1 \cdot 10^{7}$ \\ \hline
\textit{DYN5} & Chronos & subgrid dynamo & $10^{-9}$ & $84^3$ & $1024^3$ & $82$ & $8.1 \cdot 10^{7}$ \\ \hline
\textit{CSFBH2} & Chronos & \begin{tabular}[c]{@{}c@{}}cooling + chemistry +\\ star formation + AGNs\end{tabular} & $10^{-10}$ & $84^3$ & $1024^3$ & $82$ & $8.1 \cdot 10^{7}$ \\ \hline
\end{tabular}
\caption{Main parameters of the simulations analyzed in this work.}
\label{tab}
\end{table*}

\subsection{Network reconstruction}
\label{sec:net}
\begin{figure} 
    \centering
    \includegraphics[scale=0.9]{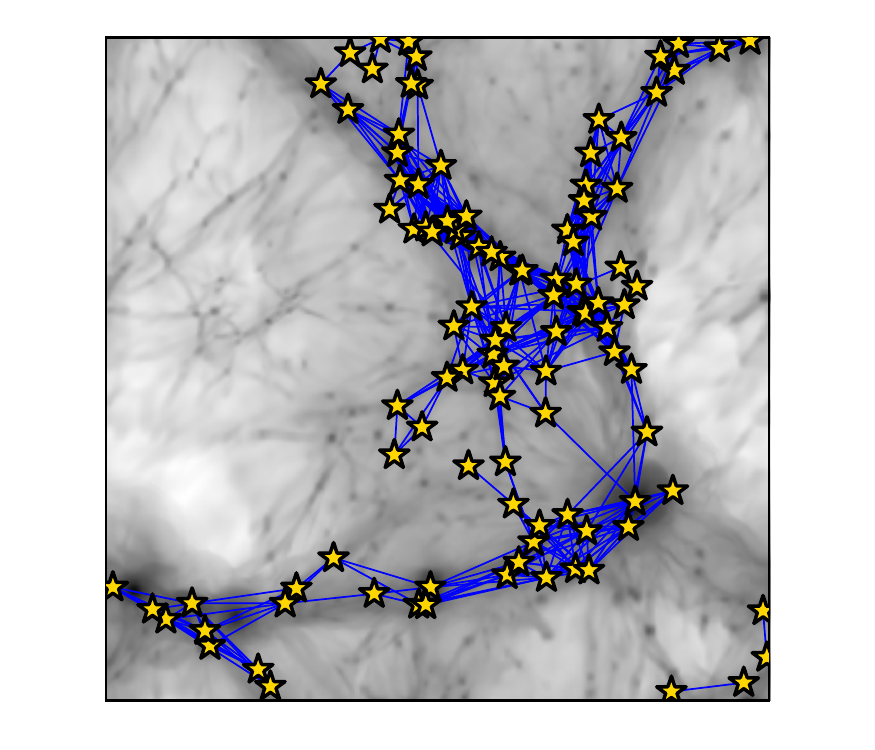}
    \caption{Projected density of a slice of $1\ \mathrm{Mpc}$ thickness and $19\ \mathrm{Mpc}$ side, with haloes and filaments obtained by our algorithm inside the selected volume.}
    \label{network}
\end{figure}

\begin{figure} 
    \centering
    \includegraphics[scale=0.9]{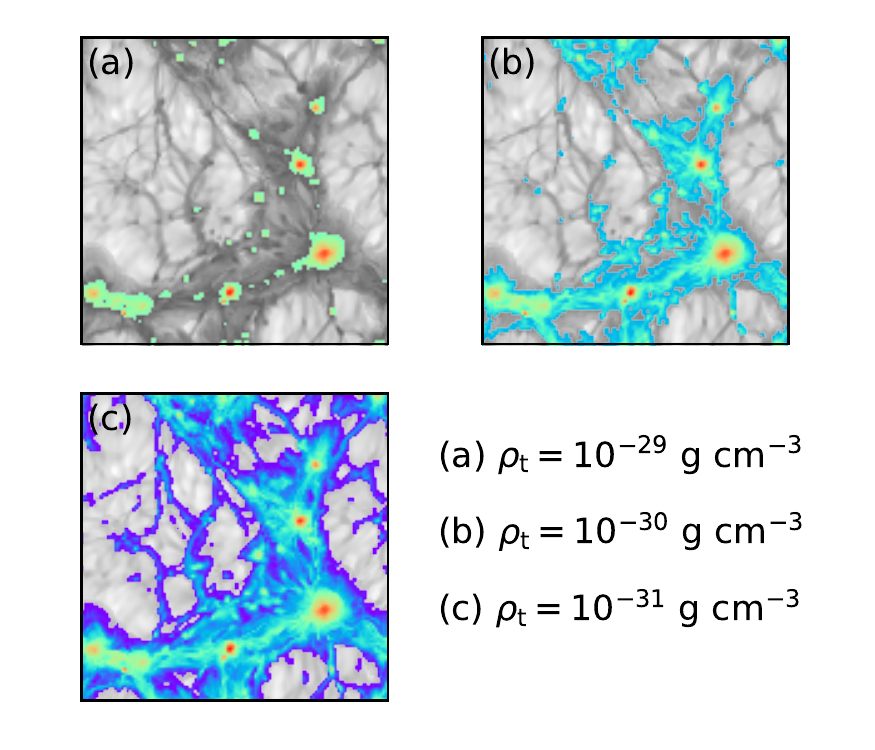}
    \caption{Gas density of a slice of $40\ \mathrm{kpc}$ thickness and $19\ \mathrm{Mpc}$ side, in greyscale (below $\rho_{\mathrm{t}}$) and in colourscale (above $\rho_{\mathrm{t}}$), for three different values of density threshold $\rho_{\mathrm{t}}=10^{-29}\ \mathrm{g\ cm^{-3}}$, $10^{-30}\ \mathrm{g\ cm^{-3}}$ and $10^{-31}\ \mathrm{g\ cm^{-3}}$. This shows that the best criterion for selecting filaments is requiring $\rho>10^{-30}\ \mathrm{g\ cm^{-3}}$.}
    \label{thresh}
\end{figure}

The network reconstruction process begins with the identification of haloes and it connects them to trace filaments. 
This simple approach has already been successfully applied to the reconstruction of the network of galaxies in real observations \citep[e.g.][]{deRegt} and also has the potential to allow comparisons with the structural properties of other natural networks \citep[e.g.][]{brain}. 
haloes are found with either a halo finding friends-of-friends (FOF) algorithm included in \textsc{Enzo} \citep[][]{enzo14} or using a halo finder developed by our group, which is more suitable to analyze large cosmological simulations \citep[e.g.][and references therein]{2020MNRAS.494.5603G}.
 haloes in the mass range $\sim 10^8 - 10^{14}\ \mathrm{M_{\odot}}$ were identified by these methods (see Table in Appendix \ref{app3} for details). Filaments are tentatively found as the line connecting two sufficiently close haloes (less than a certain distance $l_{\mathrm{c}}$ apart): if the gas density of each cell encountered by the line is above a certain threshold $\rho_{\mathrm{t}}$, then the filament is confirmed, meaning that there actually is a significant overdensity even between the two nodes.
The values of $l_{\mathrm{c}}$ and $\rho_{\mathrm{t}}$ used for our network were respectively $\sim 4\ \mathrm{Mpc}$ and $10^{-30}\ \mathrm{g\ cm^{-3}}$ for volumes of the Roger sample and $\sim 8\ \mathrm{Mpc}$ and $10^{-30}\ \mathrm{g\ cm^{-3}}$ for the Chronos suite. 
Figure \ref{network} shows the projection of a portion of the network traced by our algorithm: haloes (yellow stars) are connected by filaments (blue lines) if the density requirement is met.
The value of the density threshold was chosen by visually inspecting the selected areas for a certain range of values, as in Figure \ref{thresh}. Although even longer filaments are expected to be found in the simulated volume, we decided to narrow the sample down to filaments shorter than $4\ \mathrm{Mpc}$ or $8\ \mathrm{Mpc}$, since this criterion would allow to obtain a big enough sample, but at the same time limit the computational time required. We comment this choice in Section \ref{sec:discb}.

\subsection{Halo-filament pairing}
\label{sec:mul}
\begin{figure} 
    \centering
    \includegraphics[scale=0.9]{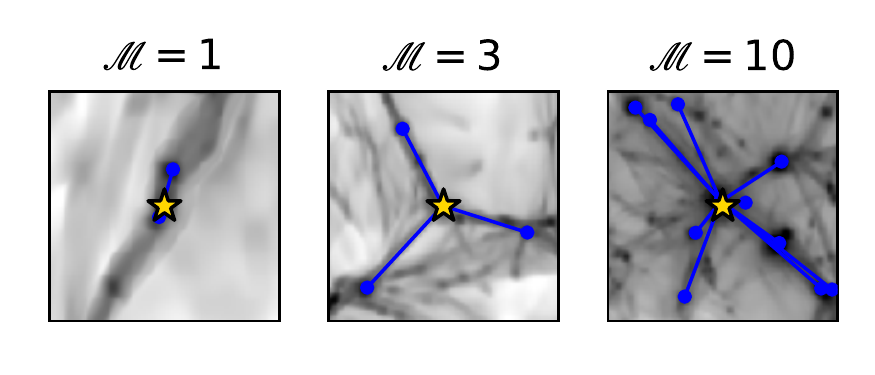}
    \caption{Examples of halo-filament(s) pairing for different multiplicities $\mathscr{M}=1,3,10$.}
    \label{netmult}
\end{figure}

Determining a correspondence between a halo and a filament is useful to find a relation between their properties, e.g. the alignment of halo spin axis and filament orientation.
To assess which filament a certain halo belongs to, we looked for filaments that connect to the halo region\footnote{We chose a volume of $\approx 400^3\ \mathrm{kpc^3}$ centred in the halo}: thus, a halo may be associated to multiple filaments, e.g. when it belongs to a cluster which connects two or more filaments. We call this property \textit{multiplicity} $\mathscr{M}$, i.e. the number of filaments corresponding to a halo (see Figure \ref{netmult}).

\section{Results}
\label{sec:res}
\subsection{The alignment of halo spin, filaments and magnetic fields}
\label{sec:rog}
\subsubsection{Network properties}
\begin{figure} 
    \centering
    \includegraphics[scale=0.9]{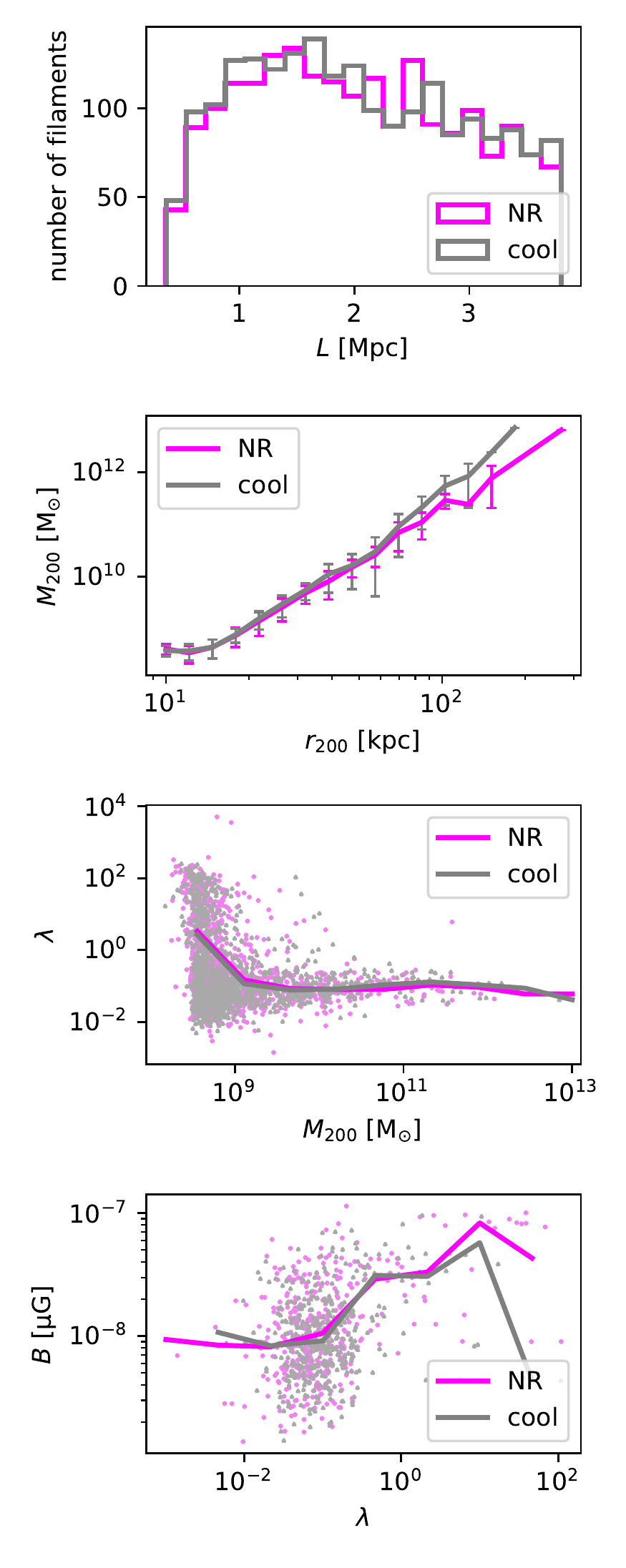}
    \caption{Comparison of halo and filament population for the NR and cool runs. First panel: distribution of filament lengths. Second panel: median virial halo mass per virial radius bin with relative error bars corresponding to the standard deviation. Third panel: scatter and median of halo spin parameter as a function of virial mass. Fourth panel: scatter and median of the magnetic field (averaged inside a $r_{200}^3$ volume around the halo) as a function of halo spin parameter.}
    \label{rcII_4}
\end{figure}

Our network reconstruction algorithm (Section \ref{sec:net}) allows to retrieve each selected filament's endpoints' coordinates inside the grid:
\begin{equation}
\mathbfit{P}_1=
    \begin{bmatrix}
    x_1 \\
    y_1 \\
    z_1
    \end{bmatrix}
\ \ \ 
\mathbfit{P}_2=
    \begin{bmatrix}
    x_2 \\
    y_2 \\
    z_2
    \end{bmatrix},
\end{equation}

such that the filament orientation follows the vector $\mathbfit{L}=\mathbfit{P}_1-\mathbfit{P}_2$ and the filament length is equal to $L=\left|\mathbfit{L}\right|$. The first panel of Figure \ref{rcII_4} shows the histograms of filament lengths: the peak is found at $\sim 1.5\ \mathrm{Mpc}$ and the distribution is mostly unaffected by cooling.

The FOF method that we used allowed to identify haloes with a total (gas and dark matter) mass larger than $\sim 10^8\ \mathrm{M_{\odot}}$. The trend of the virial mass\footnote{$M_{200}$ is defined as the total mass enclosed in a spherical volume of radius $r_{200}$, i.e. the distance from the halo centre where the average inner matter density is $200$ times the cosmological critical density.} as a function of the virial radius is shown in the second panel of Figure \ref{rcII_4}: both mass and radius have similar ranges for the two runs, but the run including cooling has a higher mass-to-radius ratio, implying a more concentrated distribution of dark matter due to baryonic infall \citep{1986ApJ...301...27B}.

The \textit{spin parameter} $\lambda$ is a measure of the rotation of a halo with respect to its potential energy \citep{1969ApJ...155..393P}. This quantity is automatically computed by \textsc{Enzo}'s halo finder according to this formula
\begin{equation}
    \lambda=\frac{J\left|E\right|^{1/2}}{GM^{5/2}},
\end{equation}
where $J$, $E$ and $M$ are the halo angular momentum, energy and mass, and $G$ is the gravitational constant.
The third panel in Figure \ref{rcII_4} gives the trend of the spin parameter as a function of halo virial mass: the evident scatter of $\lambda$ at low masses is likely an effect of the poor accuracy in the determination of the angular momentum of small haloes. Thus, in the following, we shall disregard the spin properties of haloes with $M_{200}\lesssim 10^9\ \mathrm{M_{\odot}}$. At larger masses, the curve is mostly flat, meaning that similar values of spin parameters are found in a wide range of masses. This trend in consistent with what \citet{2007MNRAS.376..215B} found for the Millennium simulation. We did not find significant changes in spin properties if cooling is turned on: this is in agreement with previous literature \cite[e.g.][]{2013MNRAS.429.3316B}.
Finally, in the bottom panel of Figure \ref{rcII_4} we show that there is a slight correlation between the halo spin parameter (considering only haloes with $M_{200}\gtrsim 10^9\ \mathrm{M_{\odot}}$) and the magnetic field strength at the corresponding location: faster rotating haloes tend to be surrounded by stronger magnetic fields, regardless of the presence of cooling mechanisms.

\begin{figure} 
    \centering
    \includegraphics[scale=0.9]{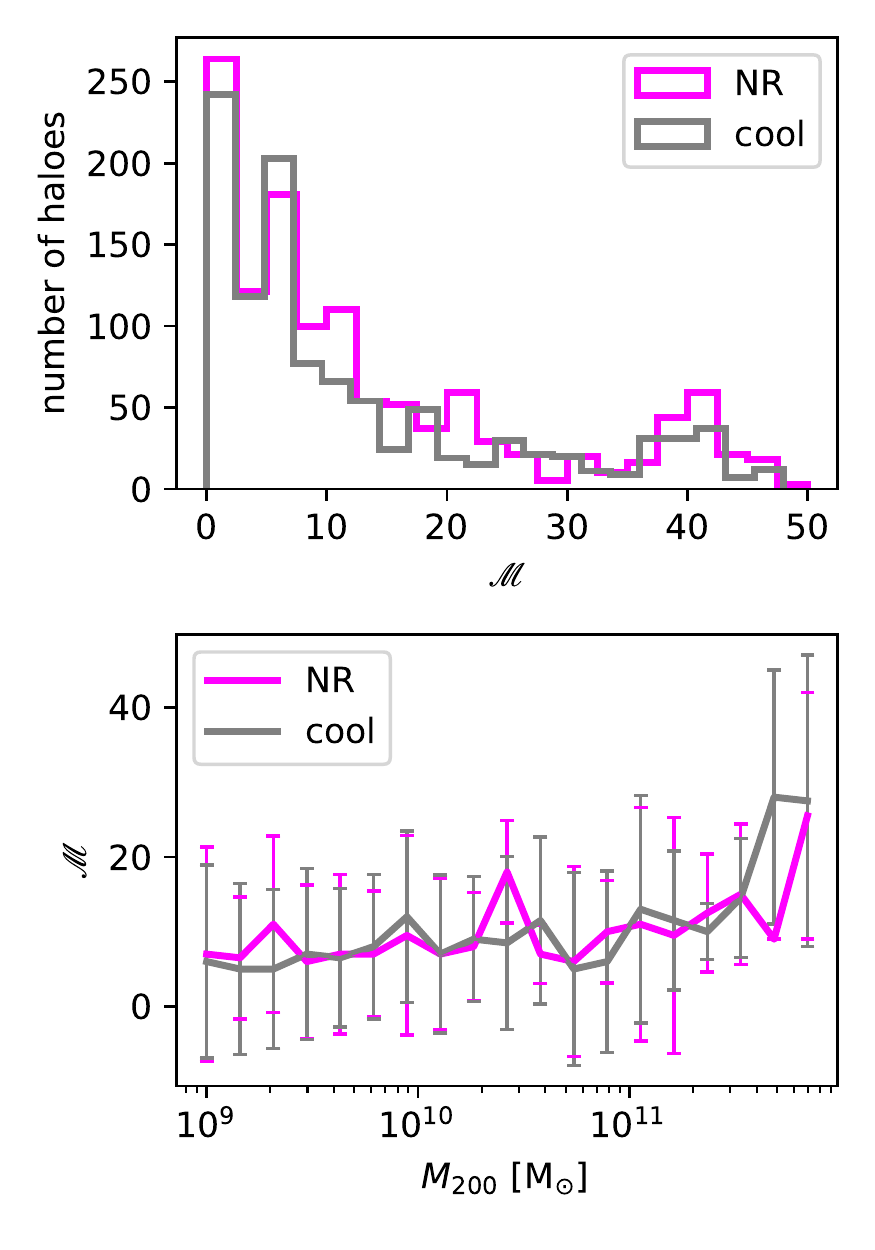}
    \caption{Distribution of multiplicity (top) and median multiplicity as a function of virial mass, with relative error bars corresponding to the standard deviation (bottom), for the NR and cool runs.}
    \label{rcII_4b}
\end{figure}

Figure \ref{rcII_4b} illustrates the properties linked to halo multiplicity: the top panel shows that most haloes are associated to a limited amount of filaments, but some of them belong to clusters, which are connected to the network through tens of filaments. Overall, cooling is not found to significantly impact on the multiplicity distribution, meaning that the number of haloes per filament (at least on the spatial scales probed by this set of simulations) is not affected by the enhanced collapse of gas structures under the effect of radiative gas cooling. In both scenarios, multiplicity correlates with halo mass (bottom panel), i.e. more massive haloes are connected to a larger number of filaments. This is consistent to what \citet{2005MNRAS.359..272C} found; also, the very high $\mathscr{M}$ values obtained for some haloes could be biased by the fact that spurious filaments are identified in very dense volumes.

\subsubsection{Spin - filament alignment}
The spin axis orientation can be obtained from the angular momentum vector $\mathbfit{J}$ of each halo, i.e. the vector sum of the angular momentum of each dark matter particle belonging to the halo.
Thus, the angle formed by the halo spin axis and the hosting filament is:
\begin{equation}
   \theta_{\mathrm{spin-filament}}=\arccos{\left(\frac{\mathbfit{J}\cdot\mathbfit{L}}{J\ L}\right)}.
\end{equation}

\begin{figure} 
    \centering
    \includegraphics[scale=0.9]{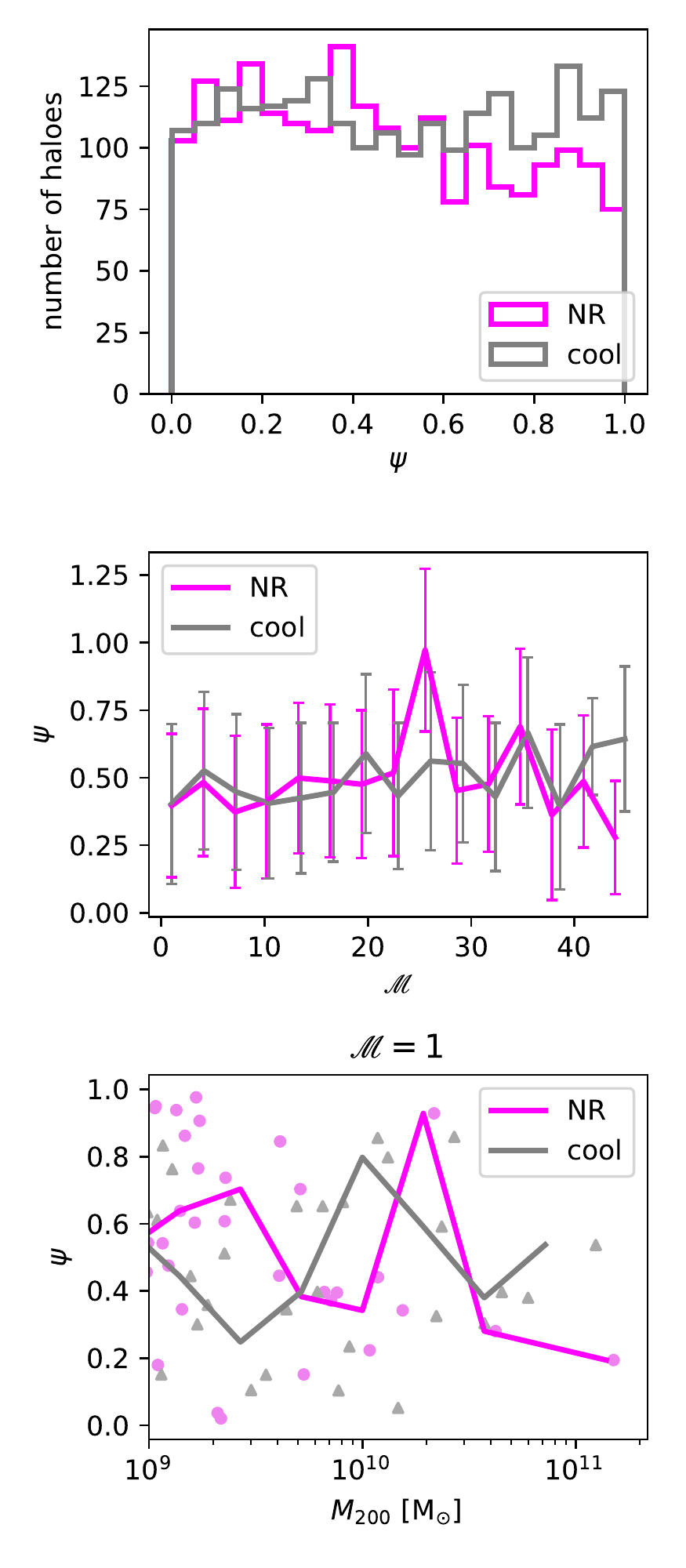}
    \caption{Comparison of halo spin behavior for the NR and cool runs. Top panel: histogram of the absolute value of the cosine of the angle between the spin direction and the host filament (if $\mathscr{M}>1$ each of the angles is included in the statistic). Central panel: median and standard deviation of the cosine of the angle between the spin direction and the host filament for multiplicity bins (if $\mathscr{M}>1$ each of the angles is included in the statistic), with relative error bars corresponding to the standard deviation. Bottom panel: scatter and median of the cosine of the angle between the spin direction and the host filament for halo mass bins if $\mathscr{M}=1$.}
    \label{fig5}
\end{figure}

This alignment is best described by the absolute value of the cosine of the angle formed by the two vectors:
\begin{equation}
    \psi_{\mathrm{spin}}=\left|\cos\theta_{\mathrm{spin-filament}}\right|,
\end{equation}

since random vectors in space form angles whose $\psi_{\mathrm{spin}}$ distribution is flat and averages to $0.5$. If a halo corresponds to multiple filaments, a value of $\psi_{\mathrm{spin}}$ is computed for each filament, i.e. $\mathscr{M}$ times.
The top panel of Figure \ref{fig5} shows the distribution of $\psi_{\mathrm{spin}}$, which is only marginally affected by the contribution of gas cooling. We can notice an excess of quasi-perpendicular configurations in the NR run, which disappears if cooling in included: a possible reason for this is that cooling enhances the accretion of denser material from filaments along more directions, which in turns tends to randomise the spin distribution.   The central panel represents the median of $\psi_{\mathrm{spin}}$ for bins of increasing multiplicity: the two curves are quite similar and there is no striking trend. In the last panel we restrict the same analysis to haloes with $\mathscr{M}=1$, in order to  study the typical behavior of halo spin in the presence of a single filament, as was done in previous works \citep[e.g. the aforementioned][]{2007A&A...474..315A}. However, the scarcity of haloes with $\mathscr{M}=1$ makes the distribution too scattered to allow us to constrain any trend, so we were not able to confirm the spin flip found in literature.

\subsubsection{Magnetic field - filament alignment}
\label{sec:magfil}
In a scenario in which cosmic magnetism is the product of primordial seed fields, cosmic structures and filaments form in a volume which is already filled by large-scale magnetic field lines. At some degree, this is also true if magnetic fields were seeded early enough for the local dynamics to rearrange the field topology.
In \citet{2020MNRAS.496.3648B}, we studied the tendency of magnetic field lines to arrange parallel to filaments' external surface during filament formation, as a consequence of shear stresses. However, while in this first work we only gave a qualitative insight of this process, here we can perform a quantitative analysis, thanks to the additional information provided by our filament reconstruction algorithm. 
In order to better analyze the alignment of a filament to the surrounding magnetic field, we first need to establish a way to trace the filaments which is more accurate than a simple straight line between two haloes: in many cases, the filament may be curved due to the presence of small haloes. Thus, the procedure to trace the actual shape of the filament is the following:
\begin{enumerate}
    \item the filament is divided into $N_r=10$ regions, whose centres are equally spaced along the filament line and whose thickness is $\sim 700\ \mathrm{kpc}$;
    \item for each of these regions, we compute the \textit{centres of mass} for the gas and use them to mark the endpoints of the $N_r-1=9$ \textit{segments} that trace each filament;
    \item the magnetic field vectors in the  $N_r-1$ cells corresponding to the segments' midpoints are then compared to the orientation of the $N_r-1$ segments. 
\end{enumerate}

\begin{figure} 
    \centering
    \includegraphics[scale=0.9]{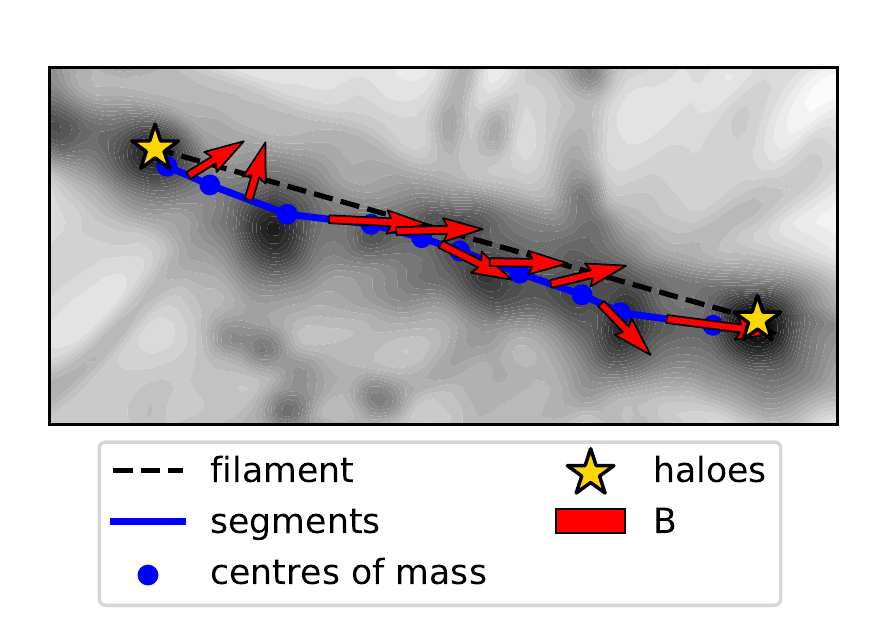}
    \caption{Projected density of a filament with local centres of mass, segments and corresponding magnetic field orientation.}
    \label{segments}
\end{figure}

Figure \ref{segments} shows a filament as an example of how the initial straight line differs from the final polygonal chain.

\begin{figure} 
    \centering
    \includegraphics[scale=0.9]{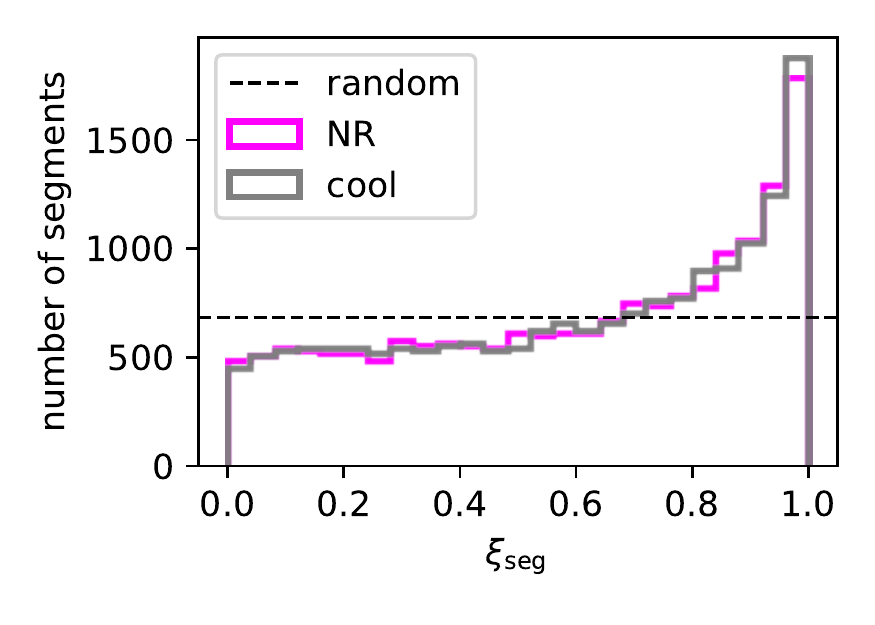}
    \includegraphics[scale=0.9]{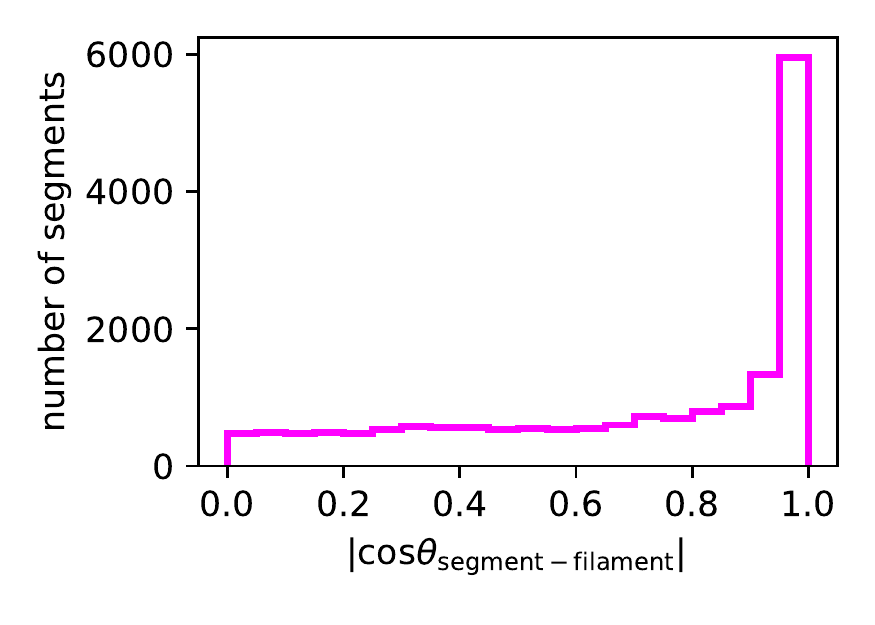}
    \caption{Top panel: histogram of the cosine of the angles formed by the magnetic field and the filaments' segments, compared to the random distribution. Bottom panel: histogram of the cosine of the angles formed by the line connecting the haloes and the segments identified by the local centres of mass.}
    \label{segfil}
\end{figure}

\begin{figure} 
    \centering
    \includegraphics[scale=0.9]{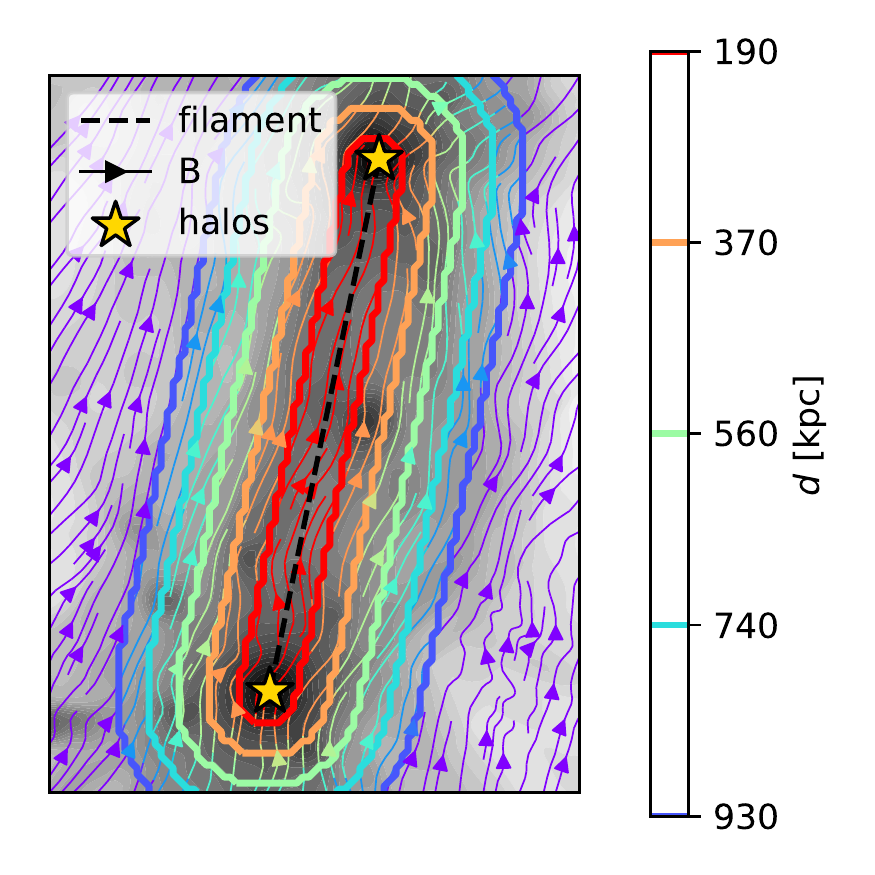}
    \caption{Projected density of a filament with contours indicating the distance from the filament with streamlines of integrated magnetic field.}
    \label{distance}
\end{figure}

The alignment of the magnetic field and the filament is parametrised by the absolute value of the cosine of the angle formed by each segment and the corresponding $\mathbfit{B}$ orientation at its midpoint\footnote{Although the value of the magnetic field at the midpoint may be subject to random fluctuations, the structures that we deal with are regular enough to ensure that no significant error is introduced, e.g. Figures \ref{segments}, \ref{distance},\ref{restest}.}:
\begin{equation}
    \xi_{\mathrm{seg}}=\left|\cos\theta_{\mathrm{B,midpoint-segment}}\right|.
\end{equation}
The top panel of Figure \ref{segfil} shows that the distribution is largely peaked at high values of $\xi_{\mathrm{seg}}$, quite distinct from the flat distribution expected for random vectors. No relevant changes are introduced by the presence of cooling mechanisms, confirming once again that the density distribution is only slightly affected.


Although the procedure involving segments is a more meticulous way to study the $\mathbfit{B}$-filament alignment, we found that the initial approximation of the filament (i.e. the line connecting two haloes) is not that far from the more accurate tracing of the filament: the bottom panel of Figure \ref{segfil} shows the distribution of the cosine of the angle formed by the initial straight line and each one of the segments composing the polygonal chain. Based on this, we can reasonably consider that the global filament orientation (at least for straight enough filaments) is sufficiently well described by the line traced by connected haloes.

Next, we want to determine the characteristic spatial scales at which the alignment develops,  i.e. how far from the filament the $\mathbfit{B}$-filament alignment is still more significant than by random chance. To do so, we consider ellipsoidal shells of gas at increasing distance from the spine of filaments, as in Figure \ref{distance}, and for each region we compute the angle formed by the magnetic field of every cell and the filament orientation.
In detail, the procedure is the following:
\begin{enumerate}
    \item for each filament, we consider a box-shaped subvolume containing it\footnote{For a filament delimited by two haloes having coordinates $(x_1,y_1,z_1)$ and $(x_2,y_2,z_2)$, the box contains all the cells that satisfy $\min(x_1,x_2)-500\ \mathrm{kpc}<x<\max(x_1,x_2)+500\ \mathrm{kpc}$, $\min(y_1,y_2)-500\ \mathrm{kpc}<y<\max(y_1,y_2)+500\ \mathrm{kpc}$ and $\min(z_1,z_2)-500\ \mathrm{kpc}<z<\max(z_1,z_2)+500\ \mathrm{kpc}$.\label{fn1}};
    \item we identify \textit{filament cells} as the ones that are intersected by the line connecting the pair of haloes (i.e. filament endpoints);
    \item for every cell in the subvolume (\textit{field cell}), the distance from each of the filament cells $d_{\mathrm{fil}}$ is computed;
    \item for each field cell, we consider the smallest distance $d_{\mathrm{min}}$ among the ones just found;
    \item we then bin the values of $d_{\mathrm{min}}$ for all the field cells, in such a way to define $5$ ellipsoidal shells and the corresponding \textit{shell cells};
    \item for each shell cell, the angle formed by the magnetic field and the filament line is parametrised by
    \begin{equation}
        \xi_{\mathrm{fil}}=\left|\cos\theta_{\mathrm{B,cell-filament}}\right|;
    \end{equation}
    \item the average of $\xi_{\mathrm{fil}}$ is computed for each shell: higher values imply a better alignment.
\end{enumerate}

\begin{figure} 
    \centering
    \includegraphics[scale=0.9]{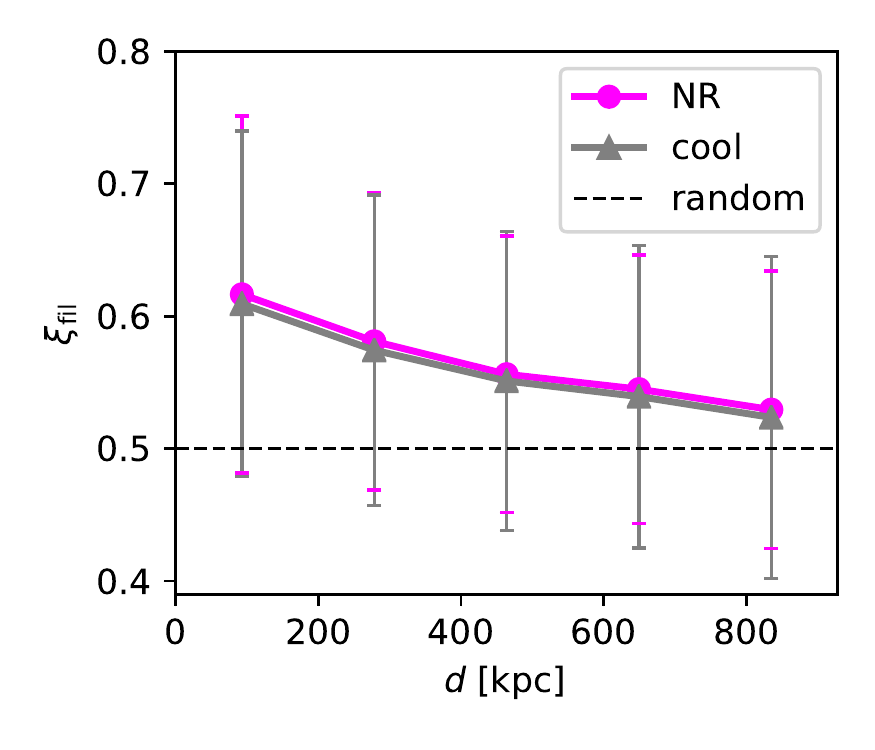}
    \caption{Median of the cosine of the angle formed by the magnetic field and the filament as a function of distance from the filament, with relative error bars corresponding to the standard deviation, for the NR and cool runs.}
    \label{kk}
\end{figure}

Figure \ref{kk} shows the median value of $\xi_{\mathrm{fil}}$ in each of the five ellipsoidal shells considered, averaged over all filaments: the magnetic field starts to align to the leading direction of filaments already at a distance of $\sim 800\ \mathrm{kpc}$ away,  and it becomes increasingly more  aligned approaching the filament spine. The presence of radiative gas cooling only moderately reduces the values of $\xi_{\mathrm{fil}}$ as a function of distance but otherwise preserves exactly the same trend: this can be ascribed to the effect of gas cooling, which tends to compress filaments towards their main axis \citep[][]{2015MNRAS.453.1164G}, hence a $\sim 10\ \%$ shift of the curve towards smaller distances.

\begin{figure} 
    \centering
    \includegraphics[clip,trim={0cm 1.7cm 0cm 0}]{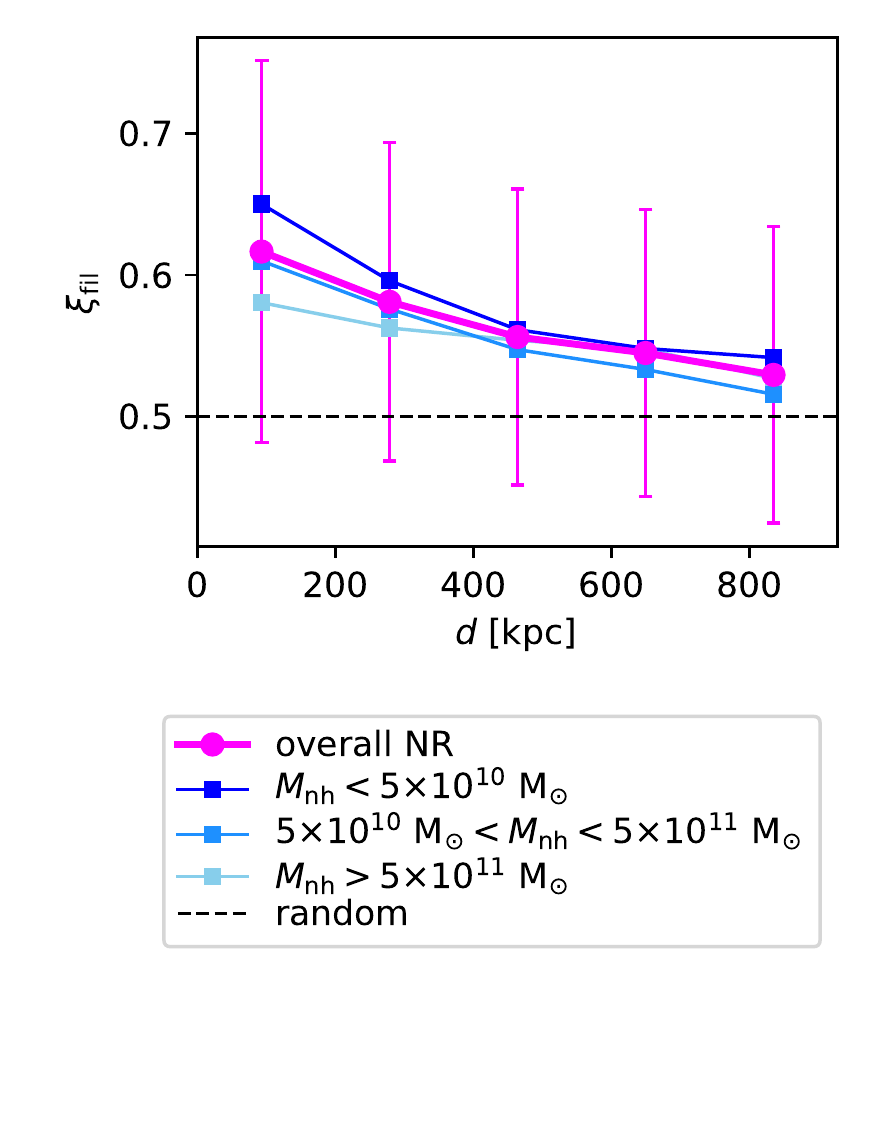}
    \caption{Median of the cosine of the angle formed by the magnetic field and the filament as a function of distance from the filament for different nearby halo masses in the NR run, with relative error bars corresponding to the standard deviation.}
    \label{ww}
\end{figure}

We also noticed that filaments with poor $\mathbfit{B}$-filament alignment typically have more haloes around them.   We quantify this property by considering the amount of haloes identified by the halo finder in the volume surrounding the filament, defined as above, weighted by their mass. In fact, a relation is found between the total mass of all nearby haloes $M_{\mathrm{nh}}$ and $\mathbfit{B}$-filament alignment (Figure \ref{ww}): the degree of alignment is significantly increased when the filament is surrounded by fewer haloes.

Incidentally, this also implies that the mass resolution of our simulations (which may affect the number of small mass haloes that can be formed in the volume) can slightly impact on the exact values of  $\xi_{\mathrm{fil}}$, since more haloes are formed for increasing resolution and thus can ``perturb'' the shape of filaments and their local alignment with magnetic fields (see Section \ref{app1}).

In summary, our preliminary analysis with a small cosmological volume, with and without the inclusion of radiative cooling, has shown that most filaments below a certain length can be described by straight lines connecting massive matter haloes, and that their shape well correlates with the topology of magnetic fields around them. In particular, we found that the magnetic field lines are well aligned to the filament both inside and outside of the overdensity, meaning that shear forces effectively drag the magnetic field, even several hundreds of $\mathrm{kpc}$ away from the accretion shocks that surrounds filaments. This means that (as extensively discussed in \citealt{2020MNRAS.496.3648B}) the alignment is not due the passage of shocks, bur rather to the global structure of the (shear) velocity field in the regions where filaments form in the hierarchical scenario. This effect is only marginally affected by non-gravitational effects, like gas cooling. 
On the other hand, the analysis of halo spin does not suggest any strong relation with the magnetic properties of the cosmic web, except for a slight tendency of magnetic fields to be stronger around haloes with higher spin parameters. Furthermore, no significant correlation between spin orientation and filamentary structures is found, unlike what previous literature suggests: this is possibly to ascribe to the limited simulated volume and resolution, and thus the small amount of haloes with reliable measures of angular momentum in our sample.

\subsection{Resolution tests}
\label{app1}
\begin{figure*} 
    \centering
    \includegraphics[scale=0.9]{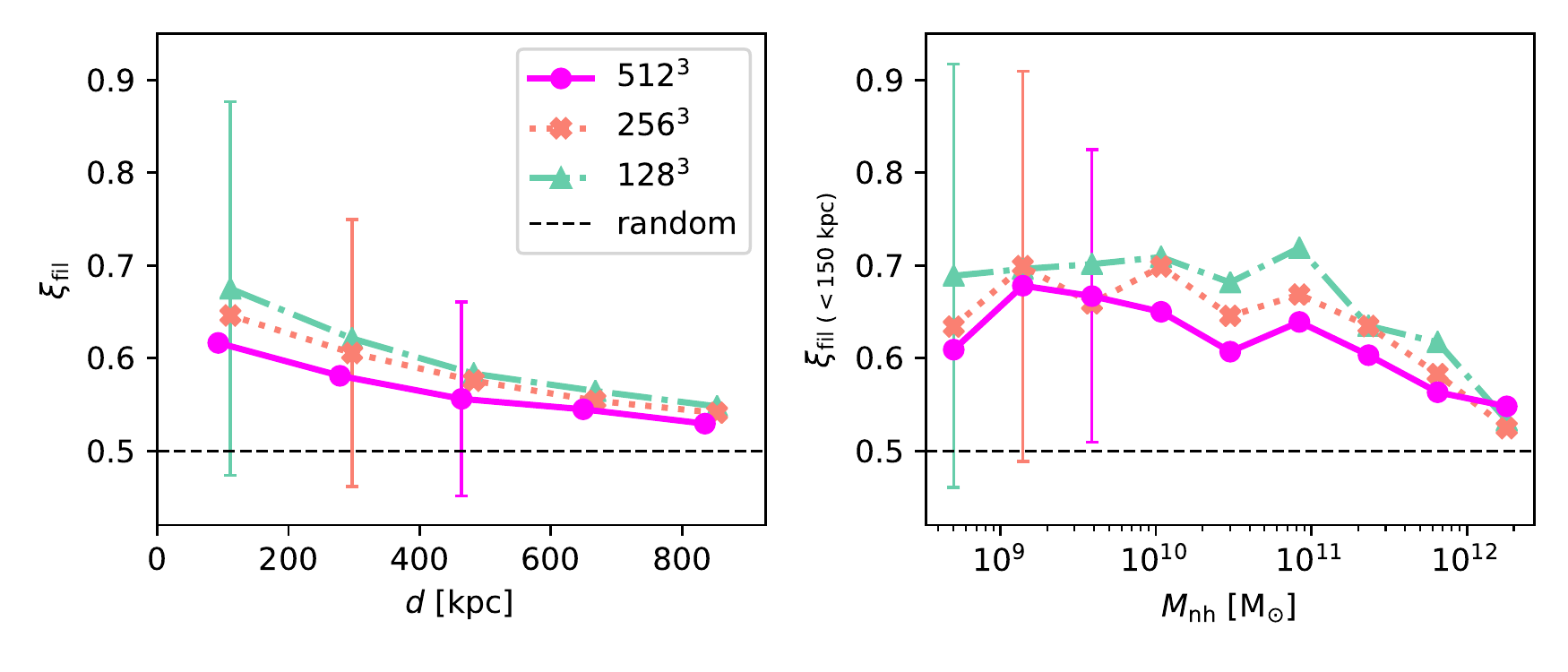}
    \includegraphics[scale=0.9]{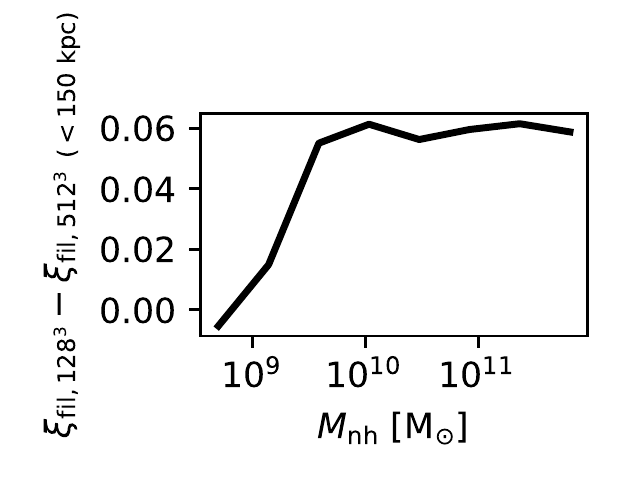}
    \includegraphics[scale=0.9]{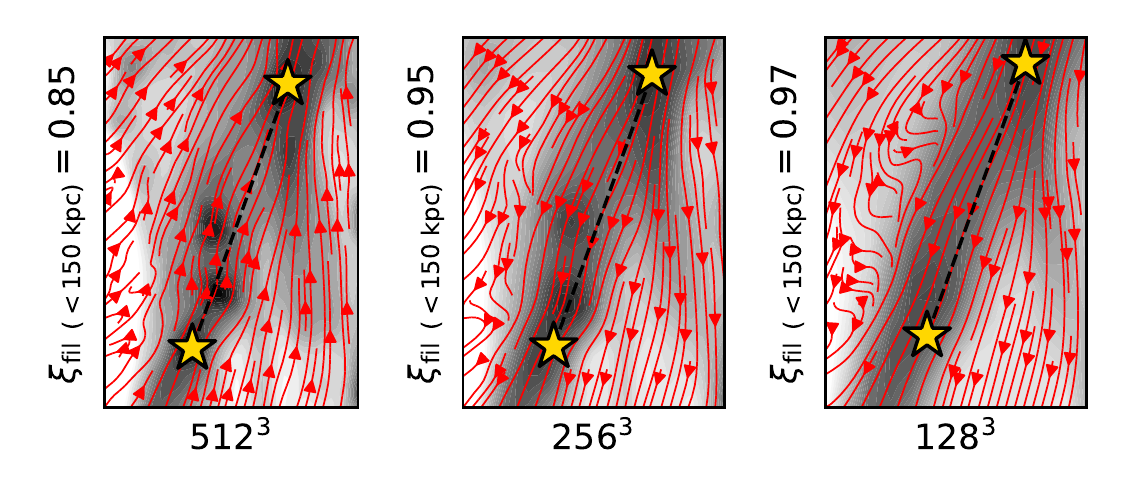}
    \includegraphics[scale=0.9]{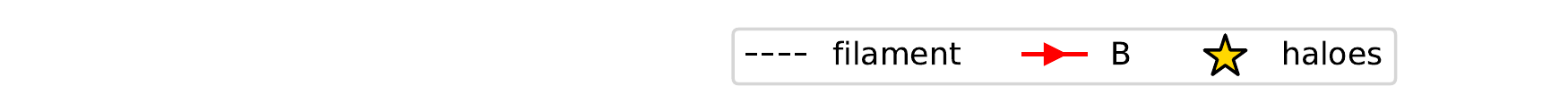}
    
    \caption{Top left panel: median of $\mathbfit{B}$-filament alignment as a function of distance from the filament for different resolutions of the NR run, with relative error bars corresponding to the standard deviation. Top right panel: median of $\mathbfit{B}$-filament alignment in cells closer than $\sim 150\ \mathrm{kpc}$ as a function of the mass of nearby haloes for different resolutions of the NR run, with relative error bars corresponding to the standard deviation. Bottom left panel: difference of the median of $\mathbfit{B}$-filament alignment in cells closer than $\sim 150\ \mathrm{kpc}$ in the $128^3$ and $512^3$ NR runs as a function of the mass of nearby haloes. Bottom right panel: projected density of a filament with integrated magnetic field streamlines for different resolutions of the NR run: at coarser resolutions smaller haloes blend with the background and the magnetic field is better aligned to the filament}
    \label{restest}
\end{figure*}

In this Section, we show the results of a resolution test on the Roger run concerning the $\mathbfit{B}$-filament alignment. We ran simulations identical to NR, except for the number of cells ($256^3$ and $128^3$, instead of the original $512^3$), so that we could compare the same simulated volume ($19^3\ \mathrm{Mpc^3}$ comoving) at different resolutions: $37\ \mathrm{kpc/cell}$, $74\ \mathrm{kpc/cell}$ and $148\ \mathrm{kpc/cell}$.
The three simulated volumes are fairly similar, so we can use the same network that we computed in the $512^3$ run: this way, filaments can be found approximately at the same location, so we can estimate the $\mathbfit{B}$-filament alignment with the new simulated magnetic field orientation and compare it to the $512^3$ run.
In particular, if we replicate Figure \ref{kk} for this set of simulations, we find higher values of $\xi_{\mathrm{fil}}$ for decreasing number of cells, i.e. better resolutions imply a slightly smaller $\mathbfit{B}$-filament alignment (see top left panel of Figure \ref{restest}). We notice that the difference is mainly relevant in the proximity of the filament, so we now focus only on the area which is less than $\sim 150\ \mathrm{kpc}$ away from the filament. 


By visually inspecting some of the filaments, we observe that the variation of $\xi_{\mathrm{fil}}$ from the higher to lower-resolution runs is more significant if haloes are found along the filament: the presence of massive structures curve the path of the magnetic field lines, lowering the $\xi_{\mathrm{fil}}$ value. This effect, however, becomes less important as the resolution worsens, since haloes are less easily formed and are blurred into the filament, thus allowing the magnetic field to proceed straight undisturbed, as in Figure \ref{restest} (bottom right panel).


To further confirm this trend, we compute, for each filament of the original $512^3$ run, the total mass of the identified haloes which can be found in the filament's surroundings, thus potentially interfering with $\xi_{\mathrm{fil}}$. We then consider the average $\mathbfit{B}$-filament alignment inside the $\sim 150\ \mathrm{kpc}$ shell and plot it as a function of the nearby haloes' total mass $M_{\mathrm{nh}}$ in the upper-right panel of Figure \ref{restest}: as expected, a better $\mathbfit{B}$-filament alignment is found where fewer haloes surround the filament. Moreover, the lower-left panel of Figure \ref{restest} shows that, if many haloes are found in the proximity of a filament in the more resolved simulation, then the difference of $\xi_{\mathrm{fil}}$ between the $512^3$ and $128^3$ run for the corresponding filament is considerable.



Nonetheless, the impact of resolution on $\xi_{\mathrm{fil}}$ is not dramatic, so we can conclude that our previous analysis is only marginally biased by our simulation's resolution. More importantly, although mass and spatial resolution may affect the absolute amplitude of the alignment in some cases, our analysis shows that the trend of $\xi_{\mathrm{fil}}$ with distance from the filament and mass of haloes are fairly robust against changes in the resolution of the simulation.

\subsection{The alignment between filaments and magnetic fields for different scenarios of magnetogenesis}
\label{sec:chronos}
\begin{figure} 
    \centering
    \includegraphics[scale=0.9]{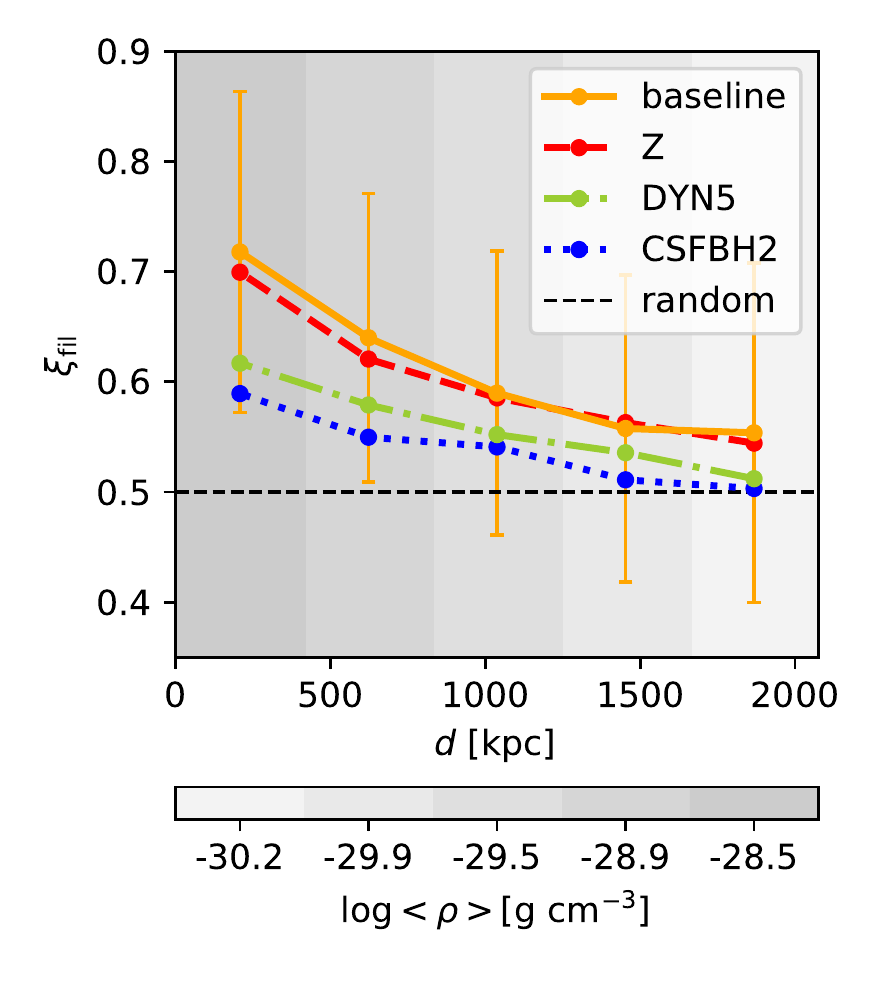}
    \caption{Median of $\mathbfit{B}$-filament alignment as a function of distance from the filament for the four Chronos runs, with relative error bars corresponding to the standard deviation. The greyscale background indicates the averaged gas density of each shell over all filaments.}
    \label{8chronos}
\end{figure}

With a second set of runs probing a much larger cosmic volume, Chronos, we tested to which extent the findings above apply to different realistic models for the origin of extragalactic magnetic fields. 
Due to the significantly larger volume and number of cells of these simulations, we perform in this case a slightly simplified analysis with respect to the one described in Section \ref{sec:rog}, i.e. we select only the most massive haloes to build the network (see the Table in Appendix \ref{app3} in the  for details) as they are the ones connected to the most prominent filaments in the simulated volume, for which we wish also to derive observational implications (Section \ref{sec:disc}). In any case, we present tests for the statistical consistency between Roger and Chronos sets, when analyzed in a similar way, in Appendix \ref{app2}.
In this Section, we focus in particular on the alignment of the magnetic field up to larger distances from filaments (Figure \ref{8chronos}). We remind the reader that we are now considering volumes $\sim 100$ times larger than we did in the previous Section: thus, working on Chronos runs, we manage to perform the analysis concerning magnetic field and filament alignment on a wider range of filament lengths (up to $\sim 8\ \mathrm{Mpc}$).

Analogously to Figure \ref{kk}, the values of $\xi_{\mathrm{fil}}$ are computed for each shell, whose typical density is indicated in grey, then averaged over all filament. The trends imply that, in runs with a strong primordial magnetic field, the alignment is enhanced and is not affected by its initial topology. On the other hand, in the simulations where no strong primordial field is present (DYN5 and CSFBH2), the alignment is less prominent, although present, especially within a few hundreds $\mathrm{kpc}$.
This confirms the scenario at which we previously hinted in \citet{2020MNRAS.496.3648B}: in DYN5 and CSFBH2 the magnetic field undergoes processes of either dynamo or magnetic feedback, which implies that it experiences a build-up over time and has not had the chance to fully align to the structures yet. On the other hand, primordial fields in the baseline and Z runs are able to adjust their orientation, following the shear motions, for a longer span of time. 

\begin{figure} 
    \centering
    \includegraphics[scale=0.9]{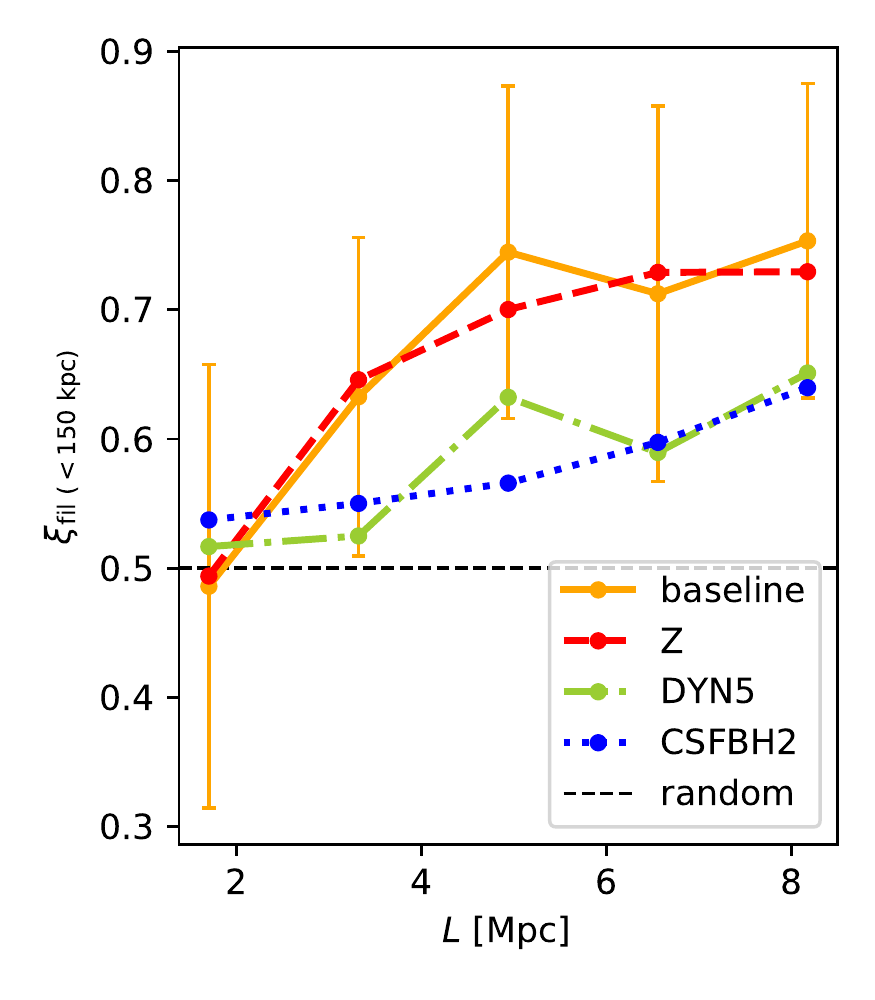}
    \caption{Median of $\mathbfit{B}$-filament alignment in the proximity of the filament (inside a $150\ \mathrm{kpc}$ shell) as a function of filament length for the four Chronos runs, with relative error bars corresponding to the standard deviation.}
    \label{anglelength3}
\end{figure}

\begin{figure} 
    \centering
    \includegraphics[scale=0.9]{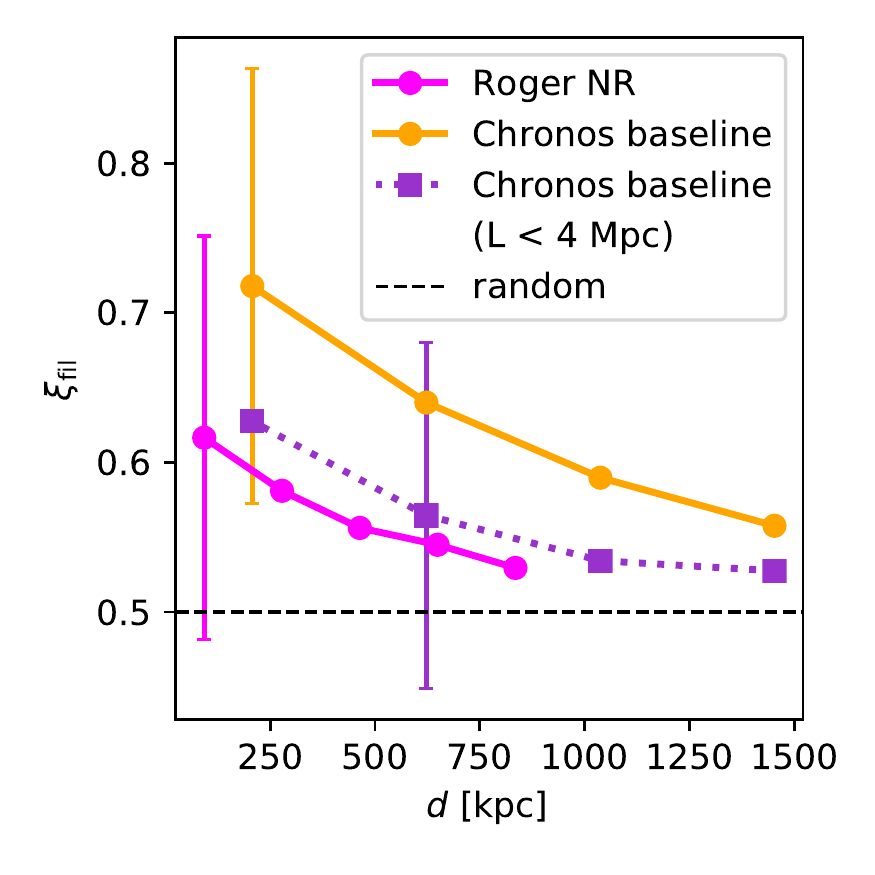}
    \caption{Median of $\mathbfit{B}$-filament alignment as a function of distance from the filament for Roger NR, Chronos baseline (for all filament lengths) and Chronos baseline's shortest filaments, with relative error bars corresponding to the standard deviation.}
    \label{anglelength2}
\end{figure}

To make sure that the results obtained from Roger and Chronos datasets are compatible, we must compare the non-radiative runs (Roger NR and Chronos baseline). Although they have similar initial conditions, there are two aspects which may cause some discrepancy in the final results: 1) the spatial resolution in the Chronos set is $\sim 2.5$ times worse than the Roger set, which would shift the curve towards higher $\xi_{\mathrm{fil}}$ with respect to the more resolved runs: however, the implications are not drastic, so this effect has a marginal impact (see Section \ref{app1}); 2) the volume simulated in Chronos is $\sim 100$ times larger, which means that there is a significantly larger population of longer filaments, which is more prone to having a better aligned magnetic field. This effect is likely to be linked to the fact that the environment around longer filaments is less perturbed by haloes at the filament endpoints, which would easily prevent the magnetic field lines from following a straight line.
In Figure \ref{anglelength3} we show how filament length is strictly related to $\mathbfit{B}$-filament alignment: that is why Chronos has, on average, higher values of $\xi_{\mathrm{fil}}$. This can be verified by comparing the $\xi_{\mathrm{fil}}$ trend as a function of distance for the same filament length range in both simulations, as in Figure \ref{anglelength2}.

To summarise, this analysis, extended to simulations of larger volumes which covered a spectrum of magnetic properites, established the role of magnetic field topology and magnetogenesis on the ability of $\mathbfit{B}$ to align to filaments, due to shear motions surrounding these structures. We can infer that this alignment is partially attenuated by the ongoing modification of magnetic fields by means of either dynamo amplification or AGN and star formation feedback.

\section{Discussion}
\label{sec:disc}
\subsection{Observational implications}
The detection of magnetic field in filaments can in principle be accomplished in two ways: through the synchrotron emission due to electrons being accelerated by magnetic fields, linked to observable radio emission \citep[e.g.][]{2017MNRAS.467.4914V,2021arXiv210109331V}, and through Faraday rotation, which rotates the linear polarization angle of the radio emission in the background, as a function of wavelength \citep[e.g.][]{2018PASJ...70R...2A}. This latter method requires the measurement of the so called rotation measure ($RM$), which is a function of the magnetic field and thermal electron density, both integrated along the line of sight \citep[e.g.][]{2002ARA&A..40..319C}:
\begin{equation}
    RM\ \mathrm{[rad/m^2]}=812\int \frac{B_{\mathrm{los}}}{\mu\mathrm{G}}\cdot \frac{n_{\mathrm{e}}}{ \mathrm{cm^3}}\cdot \frac{\mathrm{d}l}{\mathrm{kpc}}.
\end{equation}
Our work, having showed a certain tendency of magnetic field to align to filaments, suggests that it may be possible to estimate the intensity of magnetic field around filaments, starting from its line-of-sight component: in particular, if the magnetic field lines are indeed parallel to the filament, $RM$ values measured for filaments in the sky plane should highly underestimate the magnetic field intensity in that volume.
The existence of a systematic bias in the measurement of magnetic field from $RM$ implies that this technique should yield a different estimate with respect to the one inferred from radio synchrotron detection, which instead depends on the total magnetic field.
As of today, no filaments have been detected thanks to $RM$ measurements, with the exception of some excess of $RM$ signal, possibly linked to intergalactic medium \citep{2019A&A...622A..16O}.

\begin{figure*} 
    \centering
    \includegraphics[scale=0.9]{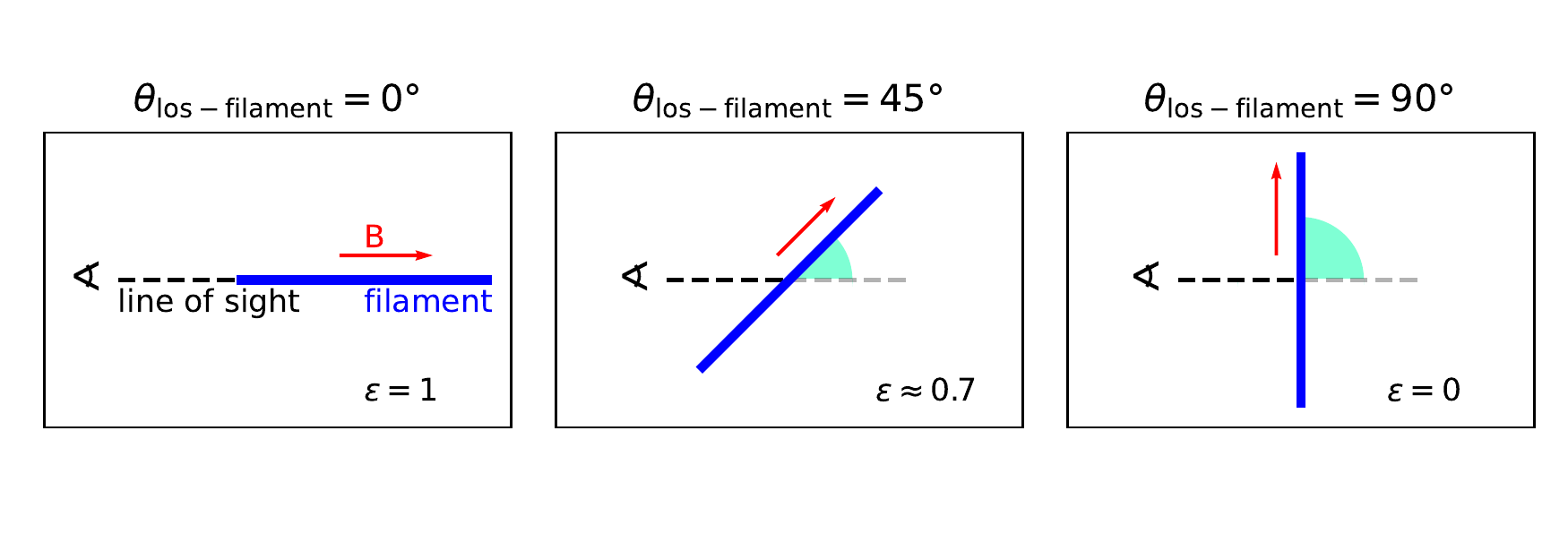}
    \caption{Schematic representation of three possible orientations of an observed filament with respect to the line of sight: $\theta_{\mathrm{los-filament}}=0^{\circ}$, $45^{\circ}$ and $90^{\circ}$. Then, assuming a uniform magnetic field perfectly aligned to the filament, the corresponding values of $\varepsilon=\left|\sum \left(B_{\mathrm{los}}\cdot\rho\right)\right|/\sum \left(B_{\mathrm{tot}}\cdot\rho\right)$ are $1$, $\approx 0.7$ and $0$.}
    \label{bf}
\end{figure*}

We measure this effect by defining the \textit{magnetic bias factor} $\varepsilon$ as
\begin{equation}
    \varepsilon=\frac{\left|\sum \left(B_{\mathrm{los}}\cdot\rho\right)\right|}{\sum \left(B_{\mathrm{tot}}\cdot\rho\right)},
\end{equation}
where the numerator contains the absolute value of the sum of the line-of-sight component of the magnetic field, weighted by each cell's density, and the denominator is the density-weighted total magnetic field.
Figure \ref{bf} illustrates three examples in which a filament is observed with an inclination of $0^{\circ}$, $45^{\circ}$ and $90^{\circ}$ with respect to the line of sight, for the simplest scenario in which the magnetic field is perfectly aligned to the filament direction throughout the whole volume. Thus, if a uniform distribution of the magnetic field were always the case, $\varepsilon$ would assume values equal to the cosine of the angle formed by the magnetic field vector and the line of sight: this distribution is flat for a random distribution of angles in space, i.e. values from $0$ to $1$ are all equiprobable, meaning that $\varepsilon$ computed for a sufficiently large sample of object would average to $\varepsilon_{\mathrm{rand}}=0.5$. However, it shall be remarked that in reality no line of sight can perfectly probe $100\ \%$ of the magnetic field, because in practice the three-dimensional distribution of magnetic fields will always fluctuate within some scale (which can change from scenario to scenario and across the variety of cosmic objects).

\begin{figure} 
    \centering
    \includegraphics[scale=0.9]{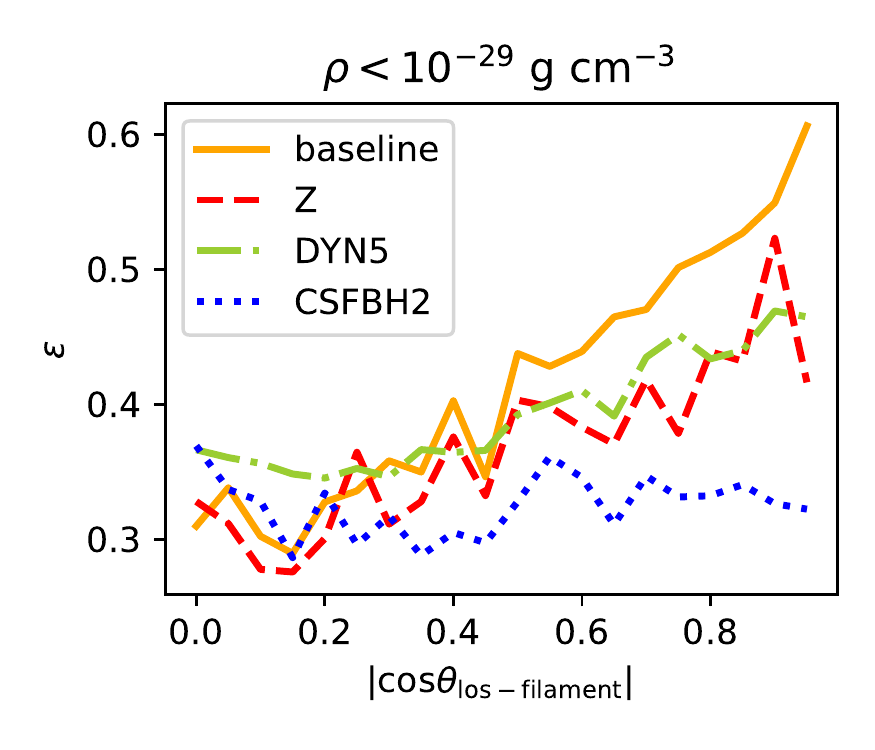}
    \caption{Median of the bias factor over all filaments as a function of the alignment between the filament and the line of sight for the four Chronos runs. The bias factor corresponding to a filament is found by performing a mass-weighted mean over the two-dimensional sky projection. Each filament is included three times, since the line of sight can be directed along any of the three coordinates. We excluded from the statistics cells in which $\rho>10^{-29}\ \mathrm{g\ cm^{-3}}$, which we assume to belong to clusters.}
    \label{obs1}
\end{figure}

We computed this value for every filament in the Chronos simulations, by considering a small volume, defined as in Section \ref{sec:magfil}, around each of them, as if they were isolated.
In order to extract the contribution of filaments alone, we computed the bias factor excluding the highest-density cells ($\rho>10^{-29}\ \mathrm{g\ cm^{-3}}$, see Figure \ref{thresh}), typically corresponding to clusters.
First, we measured the bias factor as a function of the angle formed by the filament and the line of sight, for the three spatial directions (Figure \ref{obs1}).
To better understand this plot, we note that the horizontal axis indicates the orientation of the filament with respect to the line of sight (parallel on the left-hand side, i.e. edge on, and perpendicular on the right-hand side, i.e. in the sky plane). The vertical axis contains the bias factor: lower values of $\varepsilon$ imply that the magnetic field is highly underestimated, while higher values imply that the magnetic field is less underestimated. The following particular cases correspond to specific limiting values of $\varepsilon$:
\begin{itemize}
    \item if the distribution of the magnetic field in the selected volume is completely random, then $\varepsilon=0$, since the algebraic sum of the magnetic field cancels out;
    \item if the distribution of the magnetic field is uniform in all the selected volumes, then the average over multiple objects returns $\varepsilon=\varepsilon_{\mathrm{rand}}=0.5$.
\end{itemize}

In Figure \ref{obs1}, simulations with a primordial magnetic field (baseline and Z), or with a dynamo-amplified magnetic field (DYN5) show a clear growing trend, compatible to a configuration in which magnetic fields tend to align to filaments. The values of $\varepsilon$ in CSFBH2 run, on the other hand, settle around $\sim 0.3-0.4$, meaning that the $\mathbfit{B}$-filament alignment is much more reduced in amplitude, while the randomizing effect of AGN feedback on magnetic field, around galaxies in filaments, generally decreases the average bias factor values along most lines of sight.

\begin{figure} 
    \centering
    \includegraphics[scale=0.9]{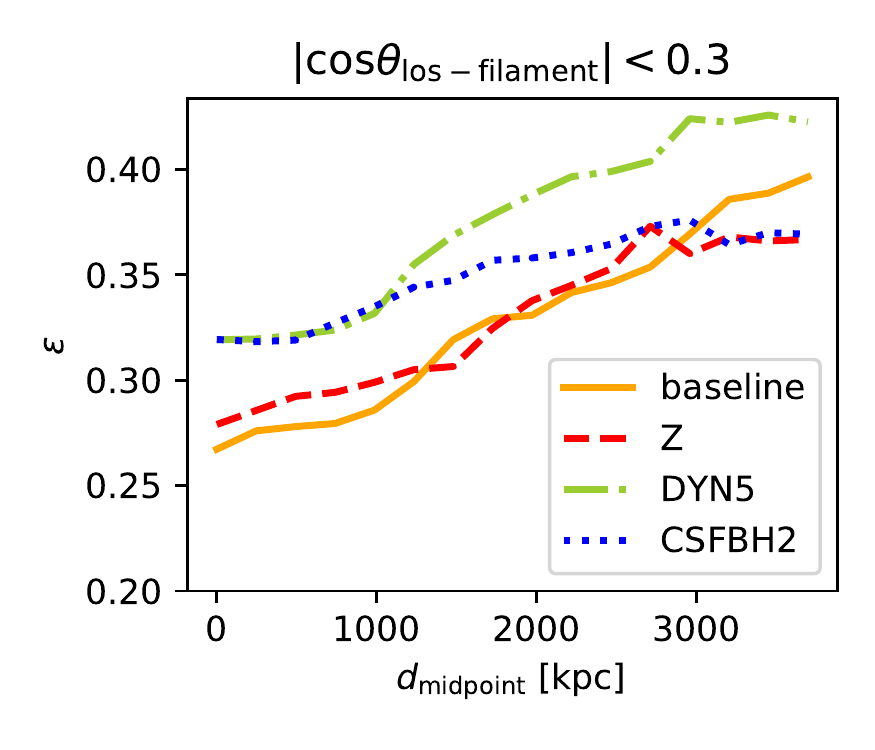}
    \caption{Median of the bias factor in the proximity of the filament (distance from the filament axis $\lesssim 400\ \mathrm{kpc}$), as a function of the distance from the filament's midpoint for the four Chronos runs. Only filaments on the sky plane, i.e. $|\cos\theta_{\mathrm{los-filament}}|<0.3$ for any of the three lines of sights, are considered. We excluded from the statistics cells in which $\rho>10^{-29}\ \mathrm{g\ cm^{-3}}$, which we assume to belong to clusters.}
    \label{obs2}
\end{figure}

We then focus on a subset of filaments roughly aligned to the plane of the sky, which are objects most suitable for observations \citep[e.g.][]{2015eheu.conf...14E,2017arXiv170905024T,2019Sci...364..981G} or stacking analysis \citep[e.g.][]{2021arXiv210109331V}. The criterion we chose for the position of the filament is that  $\left|\cos\theta_{\mathrm{los-filament}}\right|<0.3$. Figure \ref{obs2} shows the trend of the bias factor along the filament length, as a function of the distance from its midpoint. In all four simulations, $\varepsilon$ is smaller  closest to the filament's midpoint and grows as the distance increases, compatibly to the fact that, especially for the baseline and Z runs, the $\mathbfit{B}$-filament alignment is best where the filament is least affected by the clusters at the endpoints.

\begin{figure} 
    \centering
    \includegraphics[scale=0.9]{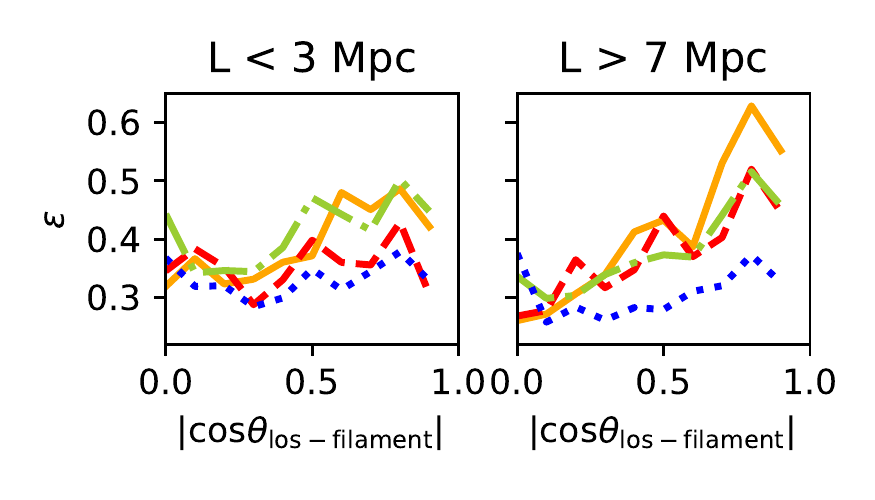}
    \caption{Same as Figure \ref{obs1}, with the additional distinction between filaments shorter than $3\ \mathrm{Mpc}$ and longer than $7\ \mathrm{Mpc}$.}
    \label{obs3}
\end{figure}

Then, we estimated the dependence of this trend as a function of the filament length: Figure \ref{obs3} replicates Figure \ref{obs1} for the $\sim 50$ longest and $\sim 50$ shortest filaments. The difference between the two cases is not large, but we notice a clearer increasing trend for the selection of longer filaments, even for the CSFBH2 run. This is reflected by a better $\mathbfit{B}$-filament alignment for longer filaments, as previously found (see Figure \ref{anglelength3}).

\begin{figure*} 
    \centering
    \includegraphics[scale=0.9]{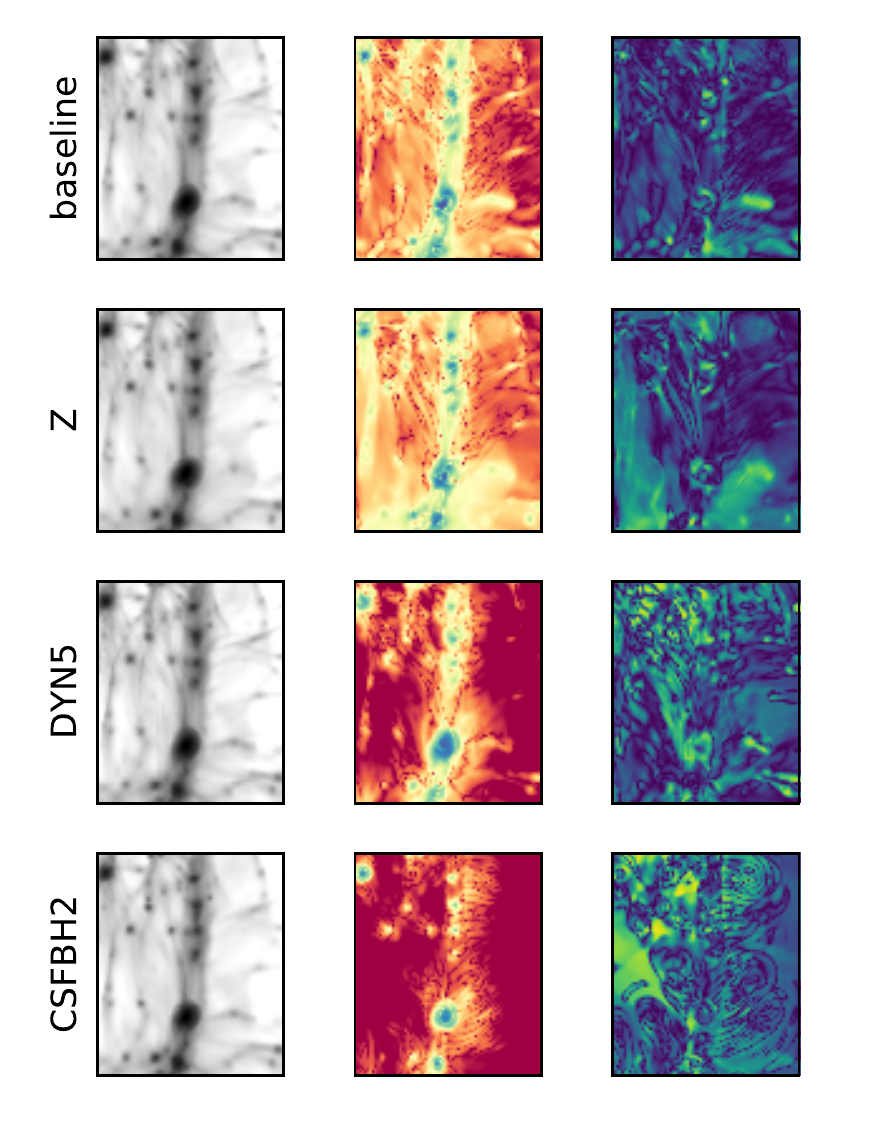}  
    \includegraphics[scale=0.9]{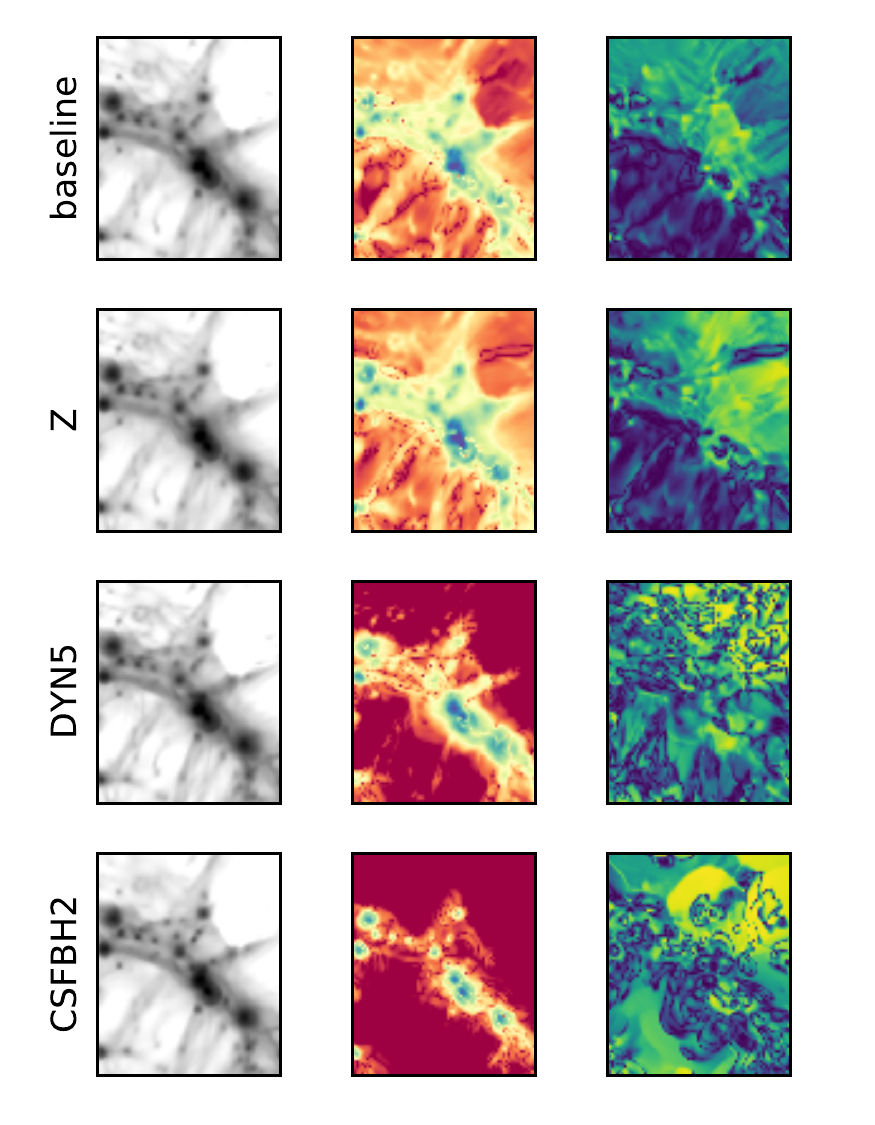} 
    \includegraphics[scale=0.9,clip,trim={0cm 0cm 0cm 15cm}]{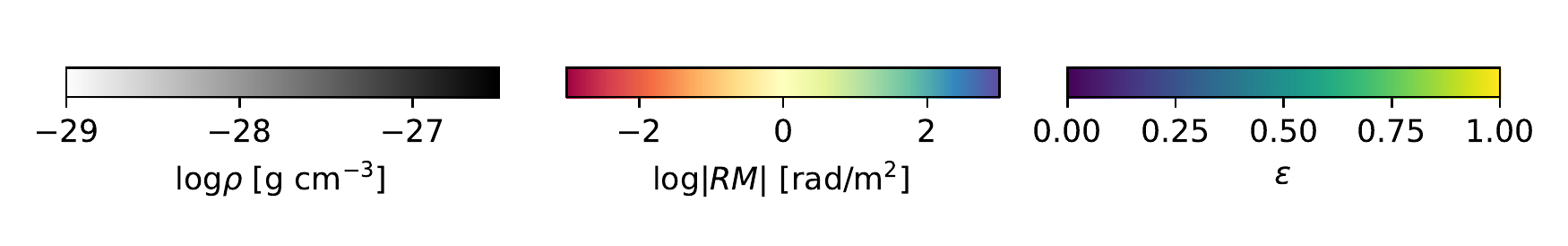}
    \includegraphics[scale=0.9]{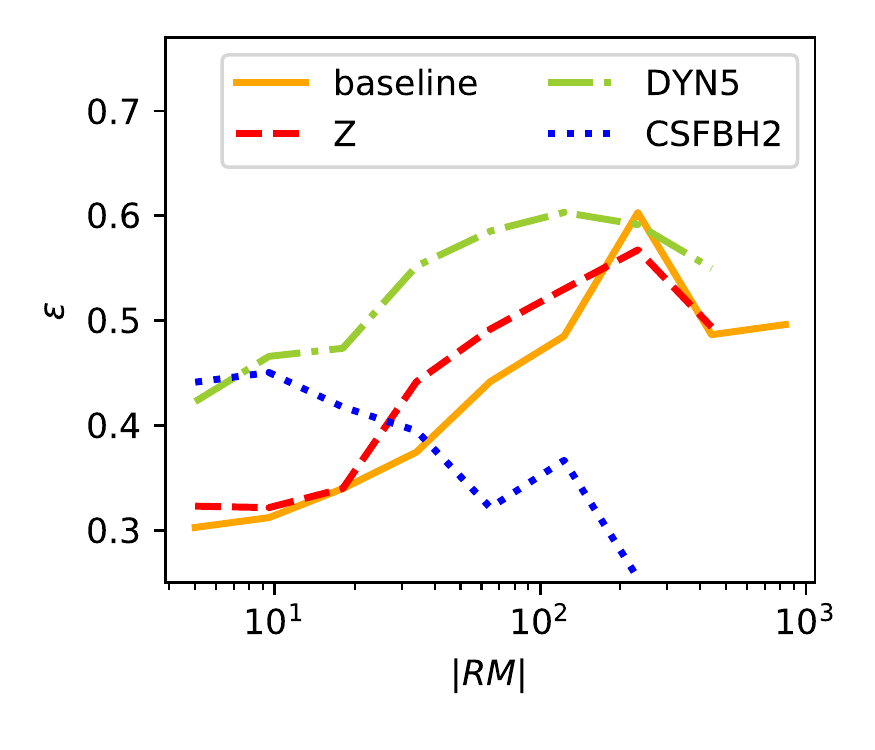}
    \includegraphics[scale=0.9]{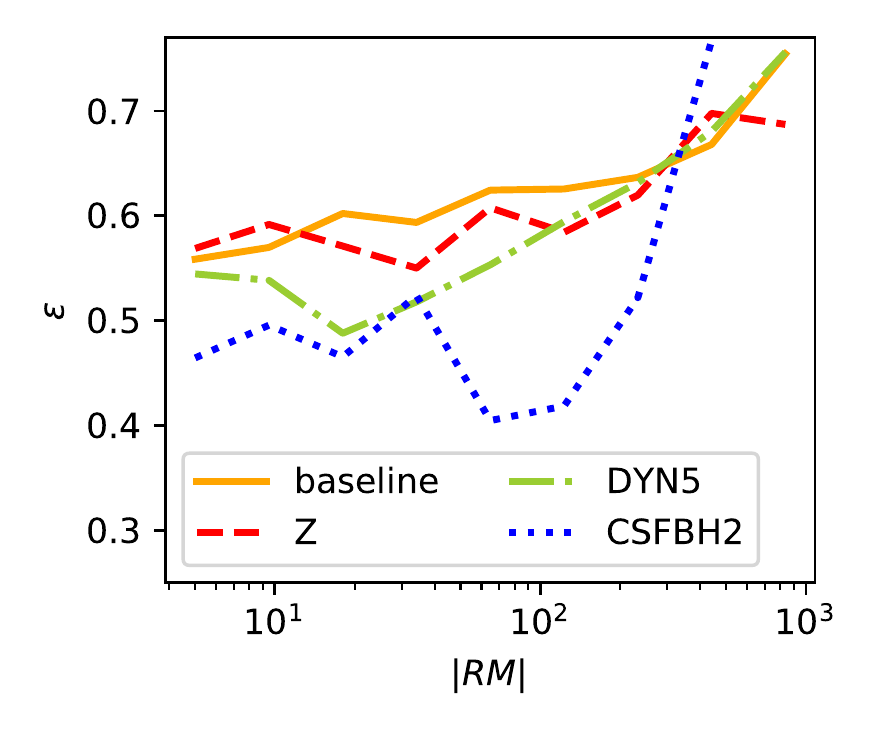}
    \caption{Top panels: projected maps of the two selected volumes containing filaments (left-hand and right-hand side) of gas density, unsigned rotation measure, and $\varepsilon$, for all Chronos simulations. Bottom panels: median values of $\varepsilon$ for bins of $|RM|$ over all cells of the two-dimensional projection for the range in which rotation measure can be detected, for the two areas represented above.}
    \label{obs4}
\end{figure*}

In Figure \ref{obs4} we show two volumes containing filaments which are almost aligned to the sky plane, as an example to illustrate the implications of this effect on the rotation measurement of such objects. The top panels show the projected maps of density, $RM$ and $\varepsilon$ for the four Chronos simulations. In presence of a large degree of alignment between magnetic fields and filaments, we therefore  expect the magnetic field to mostly lie in the sky plane as well, with a very small line-of-sight component, thereby reducing the observable  $|RM|$ towards the observer. Current instruments (e.g. VLA and LOFAR) are able to detect values of $|RM|\gtrsim 5\ \mathrm{rad/m^2}$ \citep[e.g.][]{2013MNRAS.433.3208B,2019A&A...622A..16O,2018Galax...6..128L}. The Figure suggests that, on one hand, clusters easily meet this requirement, while filaments would only be marginally detected, even for the runs in which the magnetic field is stronger (baseline and Z), due to the large degree of $\mathbfit{B}$-filament alignment, which implies low values of $\varepsilon$, as can be seen in the third column. Therefore, the small line-of-sight component yields only little  $|RM|$, typically below the detection threshold of present instruments, especially for runs in which the magnetization is weak already (DYN5 and CSFBH2). On the bottom panels we give, for each of the two selected areas, the median value of $\varepsilon$ as a function of the rotation measure (in absolute value). The highest values of $|RM|$, mostly associated to clusters, correspond to higher values of $\varepsilon$, although never approaching $\varepsilon\approx 1$; at the lower side of $|RM|$, corresponding to the areas populated by filaments, lower values of $\varepsilon$ are found, as expected from our previous considerations. The simulations are in overall agreement, except for CSFBH2, where AGN and star formation feedback introduces additional effects: although the impact of a quasi-parallel $\mathbfit{B}$-filament configuration is noticeable for a larger sample of objects (e.g. see Figure \ref{obs2}), the bursty and random occurrence of star forming/AGN events may strongly affect the local magnetic field topology and cause the statistic over a small volume to deviate from the expected trend.

The bias in the line-of-sight component typically amounts to a factor $\sim 3$ lower than the total magnetic field 
 (corresponding to $\varepsilon$ values of $\sim 0.3$).

\subsection{Numerical limitations}
\label{sec:discb}
The main numerical limitations that we encountered in this work involve the limited resolution of the simulations: we already quantified the relevance of this effect on our analysis in Section \ref{app1} and concluded that most results should be reliable and independent of resolution. On the other hand, even our post-processing algorithm for network construction is subject to limitations. For example, we restricted our analysis to filaments less than $\sim 4\ \mathrm{Mpc}$ (for the $19^3\ \mathrm{Mpc}^3$ volumes) or $\sim 8\ \mathrm{Mpc}$ (for the $84^3\ \mathrm{Mpc}^3$ volumes) long. This may seem to clash with the estimates obtained with more sophisticated network finding methods \citep[e.g.][]{2013MNRAS.429.1286C,2015MNRAS.453.1164G}, which, applied to cosmological simulations, suggest that filamentary structures up to $\sim 100\ \mathrm{Mpc}$ can form in a big enough volume. However, \citet{2015MNRAS.453.1164G}, in particular, showed that most filaments have lengths $\lesssim 10\ \%$ of the box's side length, which is compatible with our cut. On the other hand, if longer filaments were present, these would most likely be characterised by a complex morphology that would not be identified by our algorithm,  and which would make it difficult to study alignments with the surrounding haloes.  In conclusion -- bearing in mind that our goal here is manifestly not that of building a complete sample of filaments on all scales and of all possible geometries -- our method allows us to speed-up the analysis process and prevent the contamination of the sample by spurious effects, without overly limiting the statistics. Conversely, this might affect the reliability of certain inferred quantities: for example, multiplicity (Section \ref{sec:mul}) may be underestimated, since some fraction of filaments are left out.

\section{Conclusions}
\label{sec:conc}
With this work we present a simple algorithm which builds 
the network of filaments in the cosmic web in cosmological simulations, starting from the location of dark matter haloes in the cosmic volume, with the aim of producing a catalog of filaments and studying their physical properties and influence on the surrounding gas flows and magnetic fields. In particular, we looked for a relation between halo spin, filaments and magnetic field, as a function of different simulation properties, such as magnetic field initialization, presence of different astrophysical processes, and resolution. The following are our main findings:
   \begin{enumerate}
    \item morphological and dynamical features of haloes in the mass range $\sim 10^8 - 10^{14}\ \mathrm{M_{\odot}}$  (e.g. mass, spin) and filaments (e.g. length, multiplicity) are only moderately dependent on non-gravitational physics (e.g. gas cooling);
    \item in the range of lengths we considered, i.e. $\lesssim 4\ \mathrm{Mpc}$ (for the $19^3\ \mathrm{Mpc}^3$ volumes) and $\lesssim 8\ \mathrm{Mpc}$ (for the $84^3\ \mathrm{Mpc}^3$ volumes), most filaments can reasonably be  described by a straight line connecting haloes;
    \item the distribution of angles formed by magnetic field and the filament orientation in the proximity of filaments is concentrated towards quasi-parallel angles, much more than for a random three-dimensional distribution;
    \item filaments affect the shape of magnetic field lines, through the velocity shear they impose to large-scale gas flows: this effect is strongest within a few hundreds $\mathrm{kpc}$, but is still measurable down to $\sim 2\ \mathrm{Mpc}$ from the filaments' spine;
    \item the alignment between magnetic fields and filaments is particularly significant for longer filaments, which typically host fewer haloes per unit of volume; 
    \item physical models with a strong primordial magnetic field show an increased alignment between magnetic field and filaments at $z=0$,   regardless of its initial topology;
    \item weak primordial magnetic fields, later amplified by dynamo or by astrophysical processes, show less pronounced alignment, albeit still larger than in a purely random distribution;
    \item the alignment between magnetic fields and filaments is generally found to reduce the amplitude of the observable rotation measure (by a factor $\sim 3$) for filaments observed close to the plane of the sky, and it introduces a bias in the normalization of the magnetic field that can be derived from this technique.
   \end{enumerate}

To conclude, we remark that the effects above are so general (and independent on physical/numerical variations in the model) that they should also be relevant for other observational techniques probing the cosmic web, also in statistical ways \citep[][]{2021arXiv210109331V}. For example, attempts of measuring the magnetization of the intergalactic medium using fast radio bursts, which would require the combination of rotation measure and dispersion measure for the derivation of the magnetic field \citep[see][]{2016ApJ...824..105A,va18frb,2020MNRAS.498.4811H}, will
also be subject to a similar bias.

\section*{Acknowledgements}

 The cosmological simulations were performed with the \textsc{Enzo} code (\url{http://enzo-project.org}), which is the product of a collaborative effort of scientists at many universities and national laboratories. S.B and F.V. acknowledge financial support from the ERC  Starting Grant ``MAGCOW'', no. 714196. The simulations on which this work is based have been produced on Piz Daint supercomputer at CSCS-ETHZ (Lugano, Switzerland) under projects s701 and s805 and on the Marconi and Marconi100 supercluster at CINECA, under project INA17\_C4A28 and INA17\_C5A38 (with F.V. as P.I.). We also acknowledge the usage of online storage tools kindly provided by the INAF Astronomical Archive (IA2) initiative (\url{http://www.ia2.inaf.it})

\section*{Data Availability}
Relevant samples of the input simulations used in this article and of derived quantities extracted from our simulations are stored via EUDAT and can be publicly accessed through this URL: \url{https://cosmosimfrazza.myfreesites.net/scenarios-for-magnetogenesis}.



\bibliographystyle{mnras}
\bibliography{bib_align} 

\appendix
\section{Network details}
\label{app3}
In Table \ref{tab2} we indicate the details of the halo-filament network found by our algorithm in the analyzed simulations. Although the reconstruction method is essentially the same for both Roger and Chronos runs, we adjusted it before applying it to the much larger volumes involved in Chronos. This was possible due to the fact that we were no longer interested in analyzing the properties of as many haloes as possible, as we did for the Roger set in Section \ref{sec:rog}. The simplification consists of only selecting very massive haloes in Chronos runs with an overdensity algorithm and thus retrieving fewer, longer filaments. This choice allowed us to speed up the whole process on such big volumes, as well as to focus on the differences introduced by different magnetogenesis scenarios (see Section \ref{sec:detchron}). A more accurate network analysis of Chronos runs starting from the whole catalogue of haloes was performed on a small subvolume, as explained in Appendix \ref{app2}.
\begin{table}
\begin{tabular}{|c|c|c|c|c|}
\hline
 \textbf{Run} & \textbf{Number of} & \textbf{Number of} & \textbf{Maximum} & \textbf{Maximum}\\
  \textbf{} & \textbf{haloes} & \textbf{filaments} & \textbf{halo mass} & \textbf{filament length}\\ \hline

\textit{NR} & $1224$ & $1978$ & $6\cdot 10^{13}\ \mathrm{M_{\odot}}$ & $4\ \mathrm{Mpc}$  \\ \hline
\textit{cool} &  $1076$ & $2044$ & $7\cdot 10^{13}\ \mathrm{M_{\odot}}$ & $4\ \mathrm{Mpc}$ \\ \hline
\textit{baseline} &  $662$ & $226$ & $5\cdot 10^{14}\ \mathrm{M_{\odot}}$ & $4\ \mathrm{Mpc}$  \\ \hline
\textit{Z} & $662$ & $225$ & $5\cdot 10^{14}\ \mathrm{M_{\odot}}$ & $8\ \mathrm{Mpc}$  \\ \hline
\textit{DYN5} & $662$ & $223$ & $4\cdot 10^{14}\ \mathrm{M_{\odot}}$ & $8\ \mathrm{Mpc}$ \\ \hline
\textit{CSFBH2} & $662$ & $207$ & $4\cdot 10^{14}\ \mathrm{M_{\odot}}$ & $8\ \mathrm{Mpc}$ \\ \hline
\end{tabular}
\caption{Network properties in the Roger and Chronos simulations.}
\label{tab2}
\end{table}

\section{Comparison between Roger and Chronos simulations}
\label{app2}

\begin{figure*} 
    \centering
    \includegraphics[scale=0.9]{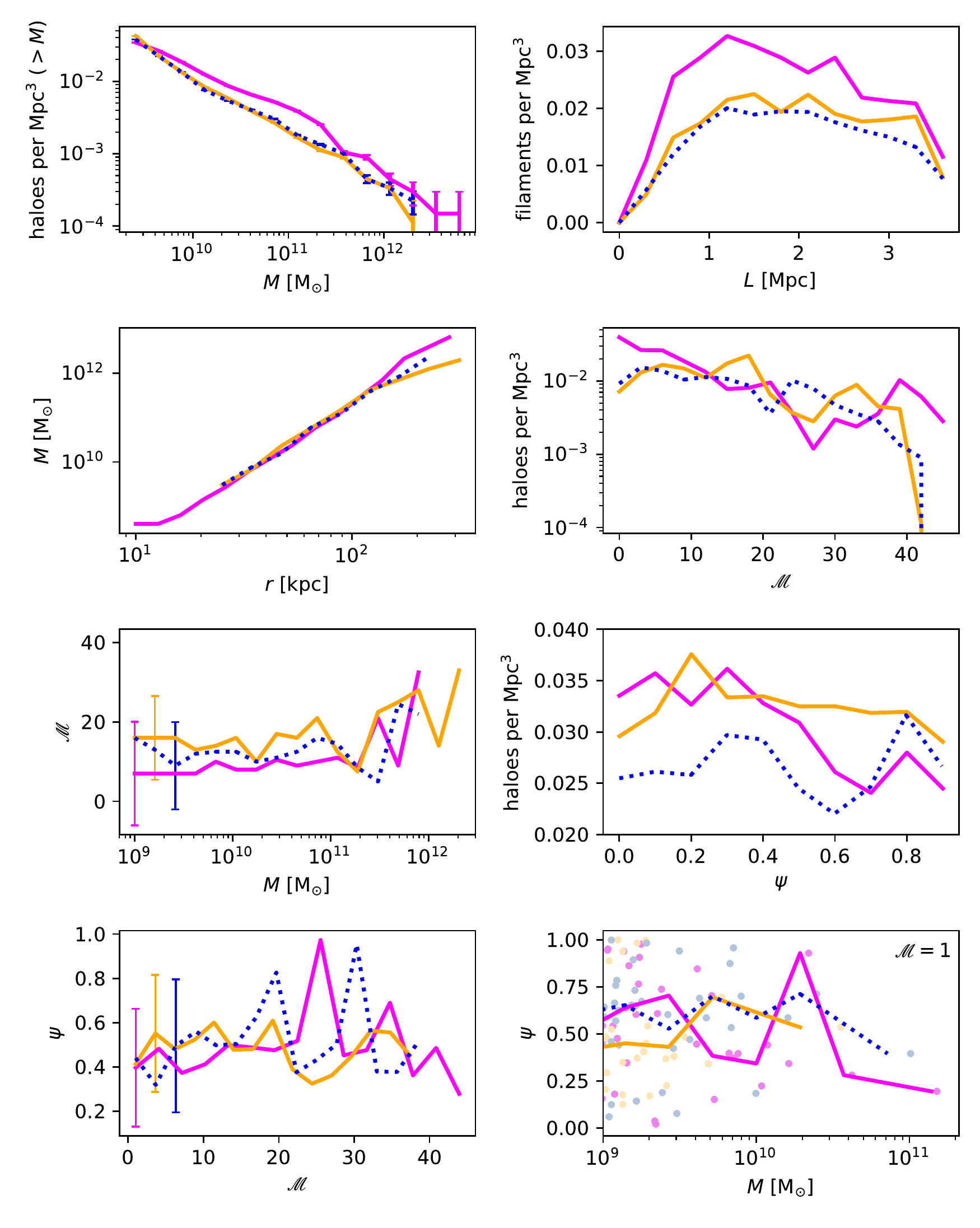}  
    \includegraphics[scale=0.9]{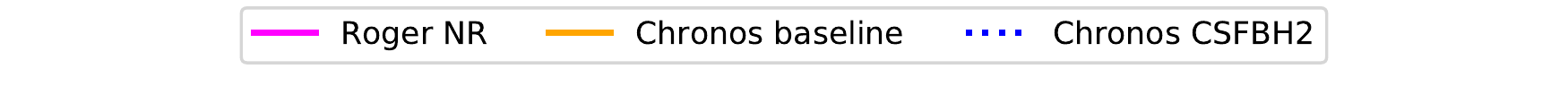}  
    \caption{Properties of haloes and filaments in the Roger non-radiative simulation and in two of the Chronos runs' subvolumes, the baseline and CSFBH2, which includes the feedback from astrophysical phenomena. From top left to bottom right: number of haloes per $\mathrm{Mpc^3}$ above a certain mass; number of filaments per $\mathrm{Mpc^3}$ as a function of filament length; median of halo mass as a function of halo radius; number of haloes per $\mathrm{Mpc^3}$ with a certain multiplicity; median of halo multiplicity as a function of halo mass; number of haloes per $\mathrm{Mpc^3}$ forming a certain spin-filament angle; median of spin-filament alignment as a function of halo multiplicity; median of spin-filament alignment as a function of halo mass for haloes with unitary multiplicity.}
    \label{figapp2}
\end{figure*}

Unlike in the main paper, for testing purposes here we apply the same network reconstruction algorithm to the cosmic web simulated in Roger and Chronos simulations (even if, in the latter case, we restrict to a $\approx 20^3\ \mathrm{Mpc}^3$ subvolume to save computing resources). In the following, we compare our most resolved $512^3$ Roger run (as in Section \ref{sec:rog}) with the baseline (non-radiative) and CSFBH2 runs of Chronos (Section \ref{sec:detchron}). 
In Figure \ref{figapp2} we reproduce some of the panels in Figures \ref{rcII_4}, \ref{rcII_4b} and \ref{fig5}: the observed trends show that, albeit with some variance related to the small volumes considered here, all main properties of haloes and filaments discussed in Section \ref{sec:rog} are also found in the considerably less resolved runs from the Chronos suite, if an identical network reconstruction is used. In any case, it shall be noticed that even if a similar mass cut in the haloes used to reconstruct the network is adopted in this case ($M \gtrsim 10^9 M_{\odot}$), the intrinsic coarser force resolution of Chronos runs leads to a $\sim\ 50\%$ reduced amount of filaments (especially shorter ones, connecting on average less massive haloes) per unit of volume. However, the dynamical correlations (or absence thereof) of gas velocity fields and the spin and multiplicity of nodes and filaments of the network, discussed in the main paper, are also confirmed by the consistent comparison of Roger and Chronos simulated volumes. In the latter case, we notice again that no relevant differences can be appreciated if radiative cooling, star formation and AGN feedback are included, once more enforcing that the main parameters of the network are not affected by these non-gravitational mechanisms.

\bsp	
\label{lastpage}
\end{document}